\documentclass[10pt,journal,compsoc]{IEEEtran}
\usepackage{graphicx}
\usepackage{capt-of}
\usepackage{amsmath,amsfonts,amssymb}
\usepackage{algorithmic}
\usepackage{algorithm}
\usepackage{array}
\usepackage[caption=false,font=normalsize,labelfont=sf,textfont=sf]{subfig}
\usepackage{textcomp}
\usepackage{stfloats}
\usepackage{url}
\usepackage{hyperref}
\usepackage{verbatim}
\usepackage{cite}
\usepackage{xspace}
\usepackage{booktabs} 
\usepackage{enumerate}
\usepackage{enumitem}
\newcommand{\name}{VibeTrace\xspace}
\renewcommand{\footnoterule}{\kern-3pt\hrule width 2in\kern 2pt}

\begin{document}

\title{From Watching to Steering: Enabling Human–Coding Agent Collaboration through Execution Traces}
\author{Xiaoke Yan, Jiaying Lei, Danqing Shi, Zichen Wang, and~Nan Cao
\IEEEcompsocitemizethanks{
\IEEEcompsocthanksitem Xiaoke Yan, Jiaying Lei, and Nan Cao are with the
Intelligent Design \& Innovation Lab, Tongji University, Shanghai, China.
E-mail: \{2433547, 2610237, nancao\}@tongji.edu.cn.
\IEEEcompsocthanksitem Danqing Shi is with the Department of Computer Science, LTH, Lund University, Sweden. E-mail: danqing.shi@cs.lth.se
\IEEEcompsocthanksitem Zichen Wang is with the School of Mathematical Sciences, Tongji
University, Shanghai, China. E-mail: sisyphusyzr@gmail.com.
\IEEEcompsocthanksitem Nan Cao is the corresponding author.
}
}

\IEEEpubid{\parbox{\textwidth}{\centering\footnotesize
This work has been submitted to the IEEE for possible publication.
Copyright may be transferred without notice, after which this version
may no longer be accessible.}}





\IEEEtitleabstractindextext{%
  \begin{center}
    \includegraphics[width=\linewidth]{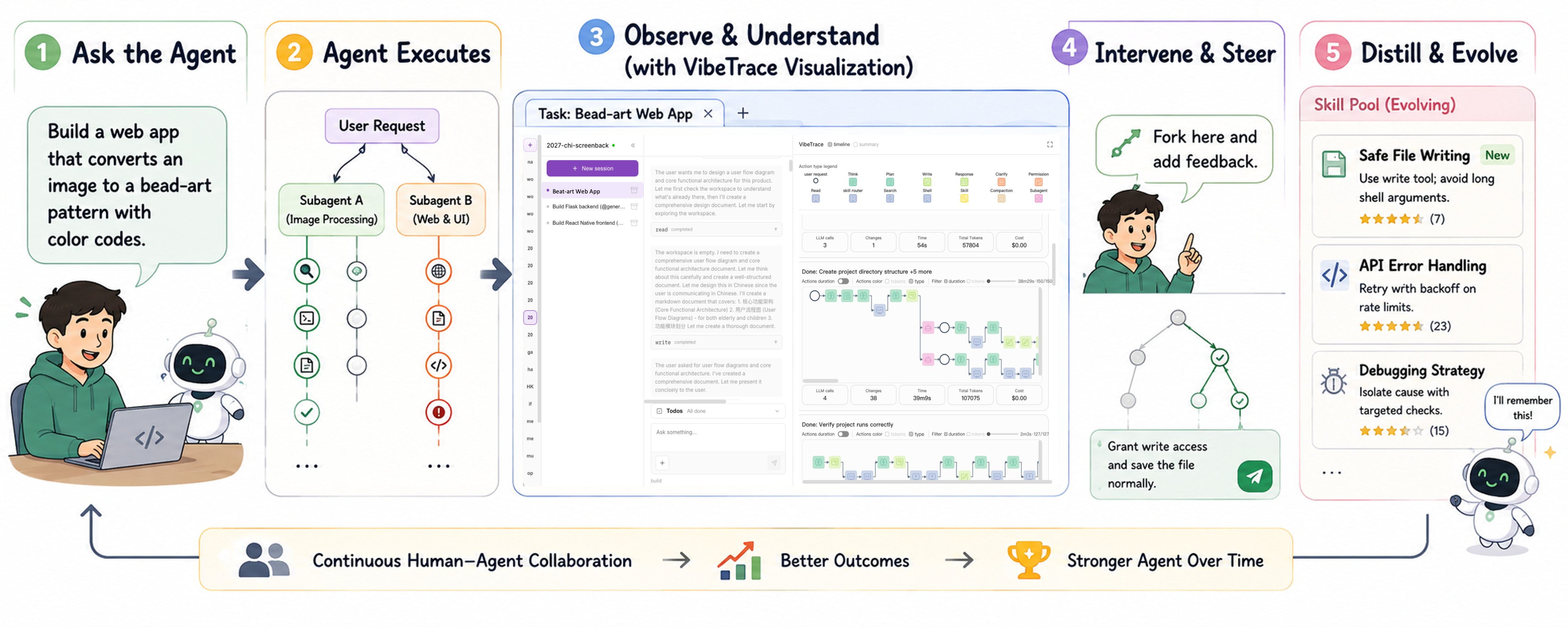}
    \setcounter{figure}{0}
    \captionof{figure}{\name enables a continuous human--coding agent collaboration loop.
      The user first assigns a coding task, after which the agent executes it while \name
      records and visualizes the execution trace. By inspecting the trace and step
      details, the user can observe and understand the agent's behavior, identify errors,
      and intervene by forking the execution and providing feedback. \name then distills
      successful human interventions into reusable skills and updates the skill pool,
      allowing the evolved agent to apply learned knowledge to subsequent tasks. This
      trace--feedback--skill loop transforms transient human interventions into persistent
      collaboration knowledge and supports continuous agent improvement. The source code is available at \textbf{\url{https://github.com/idvxlab/VibeTrace}}.}
    \label{fig:teaser}
  \end{center}
}

\maketitle

\begin{abstract}
Recent advances in large language models (LLMs) have enabled coding agents to autonomously perform complex software development tasks, yet their long-running execution remains difficult for users to inspect, understand, and steer. We present \name, an interactive trace visualization and intervention system for agentic programming built upon OpenCode. \name introduces a human--agent collaboration paradigm of \textit{Observe}, \textit{Understand}, \textit{Intervene}, and \textit{Evolve}. It (1) visualizes agent workflows as interactive execution traces for inspecting actions, intermediate states, and failures; (2) enables users to fork execution from prior steps and provide targeted corrections without restarting validated progress; and (3) distills traces and human feedback into reusable skills for continual agent evolution. On SWE-bench Verified, distilled skills reduce input tokens by 23.0\% and cost by 15.5\%, while improving pass rate from 65.0\% to 68.0\%. User studies further demonstrate improved accuracy and efficiency over conventional log-based interaction.
\end{abstract}




\section{Introduction}

\IEEEPARstart{R}{ecent} advances in large language models (LLMs) have enabled a new generation of agentic coding systems, such as Claude Code~\cite{claudecode}, OpenAI Codex~\cite{codexx}, and OpenCode~\cite{opencode}. By integrating LLMs with external tools, file systems, terminals, and execution environments, these systems can autonomously plan tasks, generate code, execute commands, debug failures, and iteratively refine solutions~\cite{yao2023react, wang2024voyager}. 
As a result, coding agents have expanded developers’ ability to accomplish complex software tasks through natural language interaction.

Despite these advances, coding agents largely follow an autonomous execution paradigm: users specify high-level goals while agents independently perform multi-step workflows. This improves efficiency, but shifts users from active problem solvers to passive observers. On complex tasks, developers must still verify the agent's direction, diagnose failures, and correct deviations~\cite{tang2023empirical, liang2024large,chen2026code}. Existing systems, however, expose mainly final outputs or linear interaction logs, offering limited support for inspection, diagnosis, and steering. Interaction is also append-only: users cannot re-enter execution at an earlier step, nor see the branch structure. Successful corrections and workflow adjustments are rarely captured as reusable knowledge, limiting improvement across tasks.

We conducted a formative study with 12 developers who regularly use Claude Code, Codex, Cursor, and OpenCode, identifying the following key for human---agent collaboration in coding tasks:
\textit{Execution transparency}: providing reliable traces that allow users to observe agent actions, decisions, and intermediate states;
\textit{Failure understanding}: revealing causal relationships between execution steps and outcomes to support effective diagnosis;
\textit{Interactive intervention}: enabling users to correct specific execution steps without restarting entire workflows;
\textit{Experience reuse}: transforming human interventions and execution traces into reusable knowledge artifacts, enabling coding agents to continuously improve through accumulated collaboration experiences.

To address these challenges, we present \name, an interactive trace-based system built upon OpenCode~\cite{opencode} that enables users to observe, understand, and steer coding agents while supporting their continual evolution. As illustrated in Fig.~\ref{fig:teaser}, \name establishes an iterative human--agent collaboration loop of \textit{Observe}, \textit{Understand}, \textit{Intervene}, and \textit{Evolve} through three key mechanisms: (1) \textit{structured trace visualization} that makes agent execution processes and dependencies inspectable; (2) \textit{trace-level intervention} that allows users to redirect execution from prior steps while preserving validated progress; and (3) \textit{online trace-to-skill distillation} that transforms execution traces and human feedback into reusable skills for future tasks.
We evaluate \name through benchmark experiments, a controlled user study, and expert interviews. Results on SWE-bench Verified~\cite{jimenez2024swebench} demonstrate improved task performance and reduced execution costs, while the controlled study shows that trace visualization improves execution understanding and error diagnosis compared to conventional logs. Expert interviews further highlight the system's usability in real-world coding tasks. The contributions are summarized as follows:

\begin{itemize}[leftmargin=0.3cm]
\item We present \textbf{\name}, an interactive trace intervention system that instantiates a new human--agent collaboration paradigm from \textbf{\textit{Observe}} to \textbf{\textit{Understand}}, \textbf{\textit{Intervene}}, and \textbf{\textit{Evolve}}. Through execution visualization, trace-level intervention, and experience reuse, \name transforms coding-agent interaction from passive monitoring of autonomous execution into active steering and continual collaboration.

\item We introduce a \textbf{visualization-oriented trace interaction paradigm} that makes agent execution processes observable and actionable. By representing agent workflows as structured execution traces, \name enables users to inspect agent trajectories, understand failures, and perform fine-grained interventions during execution rather than relying on post-hoc artifact modification.

\item We propose a \textbf{human-in-the-loop trace-to-skill distillation mechanism} that incrementally transforms execution traces and human interventions into reusable collaboration knowledge. By capturing users’ corrections, preferences, and problem-solving strategies, the mechanism enables coding agents to continuously adapt to individual workflows, improve task performance, and better align with user intentions over time.

\end{itemize}


\section{Related Work}

For background, we begin by investigating how people program with AI coding agents. Then, we review related work on agent behavior visualization, intervention, and self-evolution.
We highlight the research gap.

\subsection{Programming with AI Coding Agents}

``Vibe coding'', originally introduced by Andrej Karpathy~\cite{vibecoding}, describes an emerging programming style in which developers collaborate with AI coding agents through natural-language interaction, code generation and iterative refinement to the agent.
Such AI coding agents now support a wide range of programming tasks, including explaining code, generating tests, fixing bugs, and offering in-context code suggestions~\cite{wermelinger2023using}.
Rather than producing snippets, modern coding agents participate in multi-step developer workflows, translating high-level intent into code, synthesizing tests, and iteratively refining solutions~\cite{yao2023react, wang2024voyager}.
This shift has been shown to reshape software development practices and to improve developer efficiency and productivity, particularly for novice programmers~\cite{ziegler2024measuring}.
Prior work further shows that developers adopt different interaction strategies depending on their familiarity with the task and the complexity of the code~\cite{vaithilingam2022expectation}. With an appropriate strategy, developers can more effectively leverage AI coding agents to improve their programming experience and outcomes.

However, this new paradigm also introduces challenges related to understanding, trust, and control in AI-generated outputs~\cite{liang2024large}.
Developers often find it difficult to follow the reasoning behind AI-generated code, which complicates debugging and long-term maintenance ~\cite{tang2023empirical}. This situation highlights the need for more transparent and interpretable agents that can effectively communicate their decision-making processes to developers.
Additionally, the existing abstraction gap between users and the underlying models hampers effective usage, while limited transparency and buggy outputs can lead to frustration and decreased trust in AI coding agents~\cite{liu2023wants, wang2024investigating}.
These challenges highlight the importance of designing systems that provide developers with a better understanding and control over agent behavior~\cite{amershi2019guidelines}.
Our work builds on this area of research by exploring agent execution traces as an interaction medium, allowing users to transition from passively monitoring agent behavior to actively steering it.

\subsection{Agent Behavior Visualization and Intervention}

Visualization has long been used to make AI behavior observable, from reinforcement learning agents~\cite{wang2018dqnviz} to large language models~\cite{cheng2025understanding}.
Building on visualizations, interactive systems let users conduct inspection and shape agent behavior, for example by directly tuning reward functions~\cite{shi2024interactive}, or elicit human feedback for behavioral alignment~\cite{kompatscher2025interactive, shi2025dxhf}.
As LLM agents take on increasingly long-horizon tasks, the \emph{execution traces} the agents produce become an object of study.
One line of work makes execution traces more understandable.
Explainable-AI techniques transform raw traces into structured, human-interpretable accounts of coding-agent failures~\cite{joshi2026xai}, and multi-agent pipelines such as TraceSIR distill executions into reports that summarize what an agent did and why~\cite{yang2026tracesirmultiagentframeworkstructured}.
AgenTracer goes further by attributing an observed failure to the specific step responsible in a multi-agent system~\cite{zhang2025agentracer}.
A second line of work targets trust in what a trace reports.
VET shifts the trust anchor for agent outputs to the verifiable definition of the agent itself~\cite{grigor2026vet}. EnvTrace assesses generated code by aligning its execution against a reference run in simulation~\cite{van2025envtrace}.

The existing efforts in execution trace analysis make agent behavior more observable and easier to trust, but they treat the trace as a record to be read rather than as an ongoing process to steer.
There is still a gap remains between observing an execution and steering it, even though where and how an agent is steered strongly shapes the outcome of a run~\cite{chen2024steeringlargelanguagemodels}.
When an agent begins to deviate from the user's goal, the available way is typically to abort that execution and restart the run.
In contrast, \name treats the execution trace as an interactive artifact: users can inspect agent behavior, follow dependencies among execution steps, and intervene at the level where a problem arises rather than only at the end of the run.

\subsection{Agent Self-Evolution}

As LLM agents operate in complex environments, they increasingly require mechanisms for accumulating experience and improving future behavior across tasks. This process is often referred to as agent self-evolution~\cite{tao2024survey,fang2025comprehensive}, where past interactions are distilled into reusable knowledge that can be stored, retrieved, and refined over time. Voyager~\cite{wang2024voyager} provides an early example by continuously acquiring executable code skills from successful trajectories in Minecraft, while ReasoningBank~\cite{ouyang2026reasoningbank} extends this idea by distilling both successful and failed experiences into structured reasoning memories for test-time self-evolution.
A central form of such reusable knowledge is the agent skill. Anthropic's Agent Skills instantiate this abstraction in practice by packaging task-specific instructions, scripts, and resources into structured, retrievable modules~\cite{anthropic2025skills}. However, a static skill set is rarely sufficient: as agents encounter new tasks and failures, their skill repertoire must be continuously updated. This gives rise to skill evolution, where execution experience is transformed into long-lived and reusable skill artifacts.

Recent work explores different mechanisms for skill evolution. MUSE-Autoskill~\cite{lin2026muse} organizes skill management into a lifecycle of creation, memory, management, evaluation, and refinement. CODESKILL~\cite{li2026codeskilllearningselfevolvingskills} learns a skill-management policy for coding agents, deciding when to extract, evolve, or prune skills from trajectories. SkillOpt~\cite{yang2026skillopt} optimizes natural-language skills through trajectory-driven edits and validation-gated updates, while Skill-Pro~\cite{mi2026skill} formalizes reusable skills with activation, execution, and termination conditions and maintains them through non-parametric policy optimization. Trace2Skill~\cite{ni2026trace2skill} takes a batch distillation view, consolidating lessons from multiple execution traces into transferable skill directories. Memento-Skills~\cite{zhou2026memento} and SkillEvolver~\cite{zhang2026skillevolver} further show that externalized skills can be iteratively written, deployed, and refined without updating model parameters.

Despite this progress, most existing methods remain primarily offline or self-supervised: they rely on accumulated trajectories, model-judged outcomes, or validation feedback to update skills after execution. Recent human-in-the-loop systems begin to incorporate user feedback. AgentClick~\cite{zhuang2026agentclick} provides structured interfaces for reviewing and correcting terminal agents, and TRACE~\cite{zhou2026getting} compiles user corrections into runtime enforcement rules. However, these approaches either treat feedback as memory or enforceable constraints, rather than using interactive execution traces as direct evidence for evolving a shared skill representation. Our work addresses this gap by performing online trace-to-skill distillation during human-agent collaboration. By converting trace-level interventions, corrections, and preferences into structured skill updates, \name enables coding agents to evolve reusable skills that are temporally grounded in execution and better aligned with human intentions.

\section{Design Overview}

This section presents the design process of \name, which aims to transform coding-agent interaction from passive observation into active collaboration. We first conduct a formative study to understand how developers monitor, interpret, and intervene in agent execution during real-world coding tasks. Based on the identified findings, we derive a set of design implications that guide the design of \name. Finally, we introduce the system design, which operationalizes these principles into a trace-based human--agent collaboration framework.

\subsection{Formative Study}
To understand how users collaborate with AI coding agents beyond task completion, we conducted a formative study focusing on how users observe, diagnose, and steer agent execution in real-world software development. Our analysis revealed a fundamental gap between watching and controlling agent behavior. While users increasingly rely on execution traces to understand agent actions, they often struggle to trust process representations, diagnose failures, intervene at appropriate levels of granularity, and preserve successful intervention experiences. 

\textbf{Participants and Method.}
We conducted 40-minute semi-structured interviews with 12 experienced users of coding agents to understand how developers monitor, diagnose, and intervene in agent execution. Participants averaged 4.75 years of professional experience: eight software engineers, two product managers, and two UI/UX designers. All had used coding agents for over a year, regularly working with Claude Code, Codex, Cursor, and OpenCode.

The interview protocol focused on three stages of human-coding agent collaboration: (i) process monitoring, (ii) execution review, and (iii) intervention. Participants were asked to describe when and why they monitored agent execution, what information they relied on to evaluate agent behavior, and how they responded when execution diverged from their expectations. Follow-up questions probed recent examples, decision rationales, and breakdown situations encountered during real tasks. All interviews were transcribed and analyzed using thematic analysis~\cite{braun2006using}. Through iterative coding and constant comparison, we identified recurring themes concerning execution transparency, causal understanding, intervention practices, and knowledge reuse. These themes revealed a progression from process awareness to active steering and informed the design requirements for our trace intervention framework.

\textbf{Findings.}
Our analysis identified five interconnected findings that reveal challenges in human--coding agent collaboration and motivate a shift from passive observation to active steering and continuous adaptation.

\begin{enumerate}[leftmargin=0.5cm]
\item[{\bf F1}] {\bf Users need visibility into agent execution beyond final outcomes.}
Participants sought to understand how agents plan, act, and recover, rather than merely inspecting final artifacts. Monitoring practices varied with task complexity and confidence: most participants reviewed execution after completion, while others monitored intermediate steps for ambiguous or high-risk tasks. This highlights the need for flexible representations supporting different levels of oversight.

\item[{\bf F2}] {\bf Existing agent traces provide limited evidence for trustworthy understanding.}
Participants valued information about task decomposition, tool usage, and intermediate states but questioned the reliability of agent-generated summaries. Existing traces often describe claimed actions without verifiable evidence, making them difficult to trust. Participants therefore preferred representations grounded in observable actions and execution states.

\item[{\bf F3}] {\bf Users lack support for understanding failure causality.}
Participants found that agents could report errors but rarely explained their underlying causes. Existing logs expose isolated symptoms, such as failed commands or compilation errors, without revealing dependencies among plans, actions, intermediate states, and outcomes. Users must therefore reconstruct failure causes through additional debugging, highlighting the need for trace-based causal diagnosis.

\item[{\bf F4}] {\bf Users require fine-grained intervention to steer agent trajectories.}
Participants found existing correction mechanisms too coarse-grained when agents deviated from expectations. Continuing execution risked propagating errors, while restarting discarded validated progress. They preferred intervening at specific execution steps to redirect agent behavior while preserving successful work, highlighting the need for trajectory-level steering.

\item[{\bf F5}] {\bf Human interventions should be transformed into reusable agent skills.}
Participants frequently encountered recurring workflows, preferences, and debugging strategies but rarely retained them across tasks. Manual maintenance left valuable knowledge embedded in transient interactions, forcing users to repeat similar corrections. This motivates transforming trace-based interventions into persistent, reusable agent skills.

\end{enumerate}

\subsection{Design Implications: From Watching to Steering}
Our findings reveal a progression from observing agent behavior to actively steering and improving future collaborations. Based on this progression, we derive four design goals that establish an iterative human--agent collaboration loop: \textit{Observe} execution through trustworthy traces, \textit{Understand} behavior through causal reasoning, \textit{Intervene} by redirecting execution trajectories, and \textit{Evolve} future collaborations through reusable intervention knowledge.

\begin{enumerate}
\item[{\bf DG1}] \textbf{(Observe): Provide trustworthy and navigable representations of agent execution.}
To support monitoring and build confidence in agent behavior (F1--F2), execution should be represented through observable actions rather than agent-generated narratives. Execution traces should capture key artifacts, including task decomposition, planning, tool use, code changes, and outcomes. Given the complexity of coding workflows, traces should be organized hierarchically, allowing users to seamlessly navigate between high-level workflows and low-level execution details while maintaining situational awareness without excessive cognitive load.

\item[{\bf DG2}] \textbf{(Understand): Support causal diagnosis through trace-based reasoning.}
To help users explain execution outcomes and diagnose failures (F3), the system should expose causal relationships among actions, intermediate states, tool calls, and outcomes rather than isolated events. Making these dependencies explicit enables users to understand why failures occur, verify agent reasoning, and identify intervention opportunities.

\item[{\bf DG3}] \textbf{(Intervene): Enable trajectory-level steering through localized trace intervention.}
To support corrective actions without discarding successful progress (F4), execution traces should serve as interactive control surfaces rather than immutable records. Users should be able to inspect, modify, replace, or reject individual actions and immediately understand their downstream effects, enabling precise trajectory-level steering while preserving validated execution.

\item[{\bf DG4}] \textbf{(Evolve): Enable coding agents to improve through execution traces and human interventions.}
Coding tasks often involve recurring workflows and failure patterns, yet current agents rarely learn from prior interactions (F5). The system should transform human interventions into reusable skills that capture user preferences and effective strategies, enabling agents to continuously improve future executions, avoid repeated mistakes, and better align with user expectations.

\end{enumerate}

\subsection{System Design}

\begin{figure*}[t]
\centering
\includegraphics[width=\linewidth]{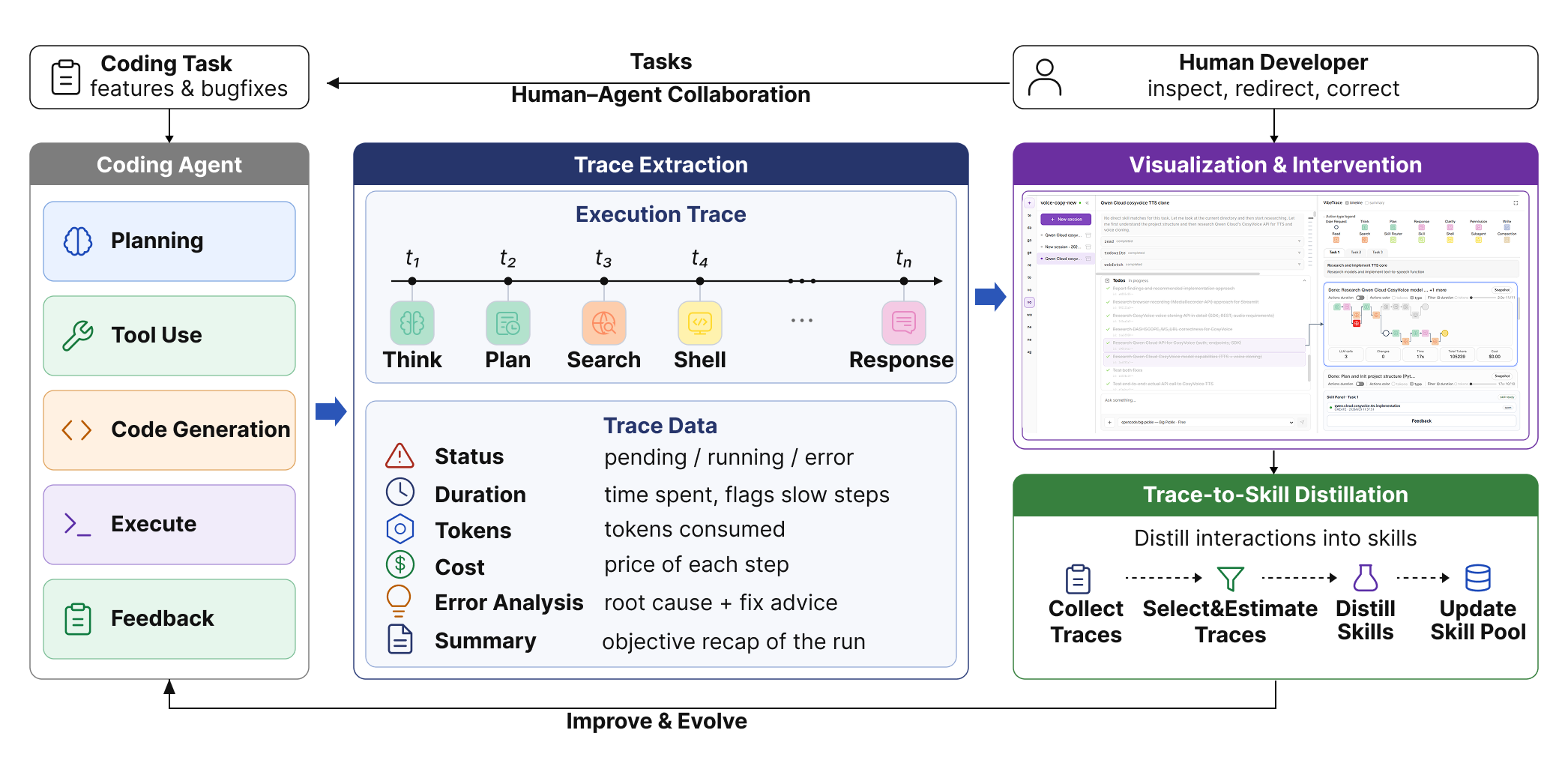}
\caption{System framework and running pipeline. \name transforms opaque agent execution into interactive traces that support workflow inspection, fine-grained intervention, and continuous skill evolution.}
\label{fig:system}
\vspace{-1.0em}
\end{figure*}

To achieve the above goals, we designed the \name system by following the architecture shown in Fig.~\ref{fig:system}. 
\name introduces a trace-based human--agent collaboration framework that transforms coding-agent execution from an opaque process into an observable, steerable, and continuously evolving workflow. 
The system consists of three core components: (1) Trace Extraction module, which captures agent execution as structured traces, enabling users to observe agent workflows and understand execution states (DG1, DG2); (2) Interactive Trace Visualization and Intervention module, which transforms traces into interactive representations that support inspection, diagnosis, and fine-grained trajectory steering during execution (DG2, DG3); and (3) Online Trace-to-Skill Distillation module, which converts execution traces and human interventions into reusable skills, allowing agents to continuously improve through accumulated collaboration experiences (DG4). In the following section, we present the technical details of each module.

\section{Technique Details}

To enable the transition from observing agent execution to actively steering and improving agent behaviors, we further detail the underlying techniques of \name. This section introduces the trace construction mechanism, interactive trace-based intervention approach, and online trace-to-skill distillation algorithm that together support human-guided coding agent evolution.

\subsection{Execution Trace Extraction}
\label{sec:extraction}
The Trace Extraction module transforms coding-agent execution into structured traces that capture the complete workflow beyond final outputs. 
During task execution, \name records agent activities, including planning, information acquisition, code modification, user interaction, skill utilization, and execution management, as a unified trace representation. 
Rather than treating execution logs as isolated events, \name organizes agent behaviors into structured trace units that preserve temporal order, intermediate states, and execution dependencies. Specifically, we define an execution trace as a sequence of trace events:
\begin{equation}
    \tau = \{e_1, e_2, ..., e_n\}, 
    \quad e_i = (a_i, s_i, o_i, t_i, r_i)
\end{equation}
where $a_i$ represents the executed action, $s_i$ denotes the corresponding execution state, $o_i$ captures the generated output or intermediate artifact, $t_i$ records the execution time, and $r_i$ represents the resulting state after the action.

Based on the execution schema of OpenCode, \name extracts 12 primitive actions grouped into seven categories. 
The first category includes \textit{\textbf{planning actions}}, which capture agent planning and decision-making processes through LLM invocation. 
The second category contains \textit{\textbf{information acquisition actions}}, including reading and searching operations that collect contextual information from files, code repositories, or external resources. 
The third category consists of \textit{\textbf{code modification actions}}, which record writing operations that update source code or other development artifacts. 
The fourth category includes \textit{\textbf{communication actions}}, including responding, clarifying, and permission requests, where the agent communicates intermediate results or involves users to resolve ambiguity and obtain authorization. 
The fifth category contains \textit{\textbf{skill management actions}}, including skill routing and skill loading, which record the retrieval and application of reusable procedural knowledge during execution. 
The sixth category includes \textit{\textbf{execution actions}}, such as shell execution and subagent invocation, representing external process execution and task delegation. 
The final category is \textit{\textbf{context management actions}}, such as compaction operations that compress execution history when context limitations are reached.


\begin{figure*}[t]
\centering
\includegraphics[width=\linewidth]{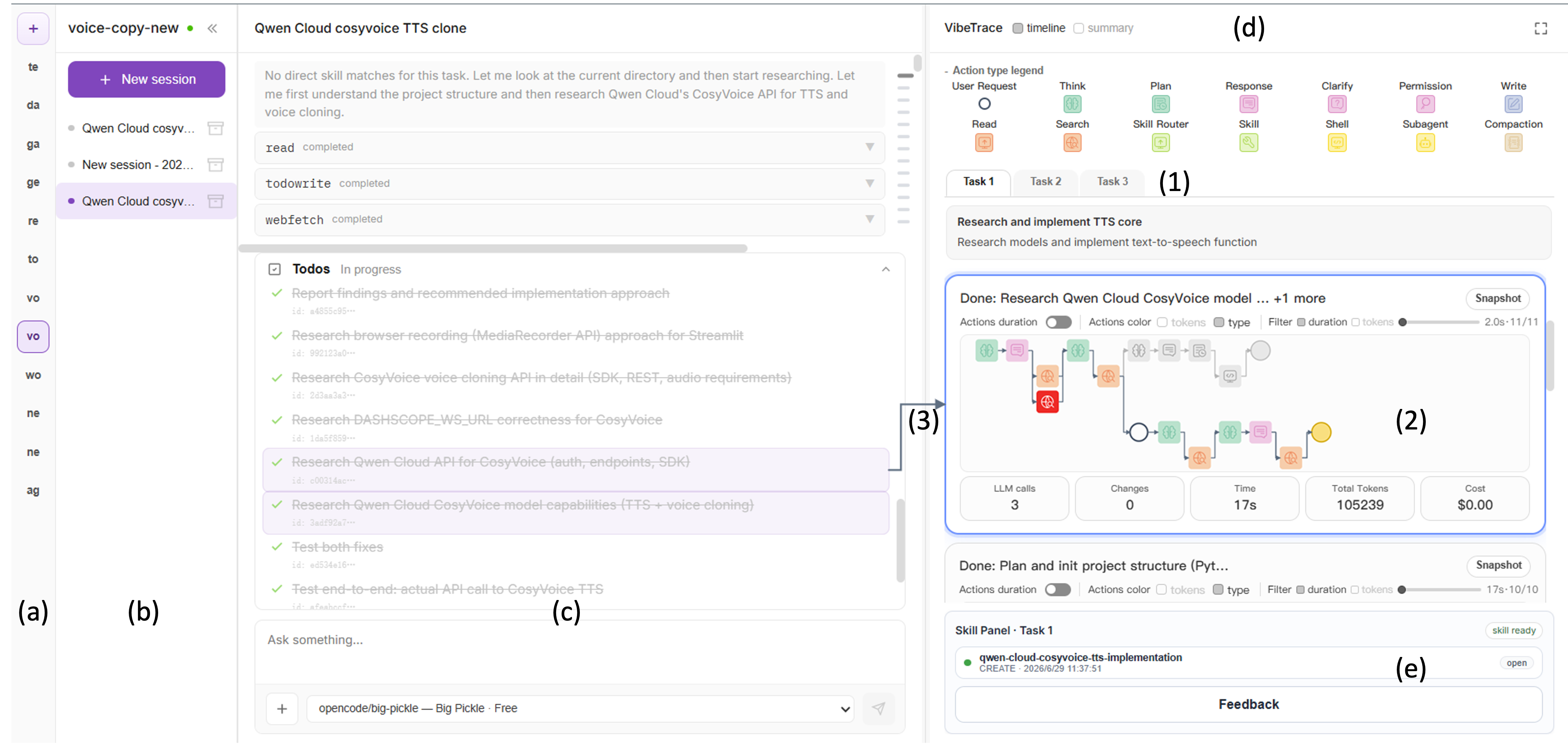}
\vspace{-1.0em}
\caption{Interactive user interface.}
\label{fig:ui}
\vspace{-1.0em}
\end{figure*}

\subsection{Interactive User Interface}
Fig.~\ref{fig:ui} illustrates the interactive interface of \name. Unlike conventional chat-based coding agents that expose execution as linear conversation logs, \name treats execution traces as the primary interaction medium through which users observe, understand, and steer agent behaviors. The interface comprises five coordinated views: a \textit{Project Panel} for project management (Fig.~\ref{fig:ui}(a)), a \textit{Session Panel} for interaction history (Fig.~\ref{fig:ui}(b)), a \textit{Dialog Panel} for natural-language communication (Fig.~\ref{fig:ui}(c)), a \textit{Trace Panel} for trace representation and interaction (Fig.~\ref{fig:ui}(d)), and a \textit{Skill Panel} for displaying reusable skills distilled from previous executions (Fig.~\ref{fig:ui}(e)). Among them, the Trace Panel serves as the primary workspace where users inspect and steer agent execution based on an interactive trace visualization which is detailed in the next.


\subsubsection{\bf Visual Trace Representation}
To improve execution transparency (DG1) and facilitate users' understanding of agent behaviors (DG2), \name organizes execution traces hierarchically, enabling users to progressively inspect agent execution from task-level workflows to individual execution actions.

\textbf{Task-level Organization.}
Each coding task is presented as an independent tab (Fig.~\ref{fig:ui}(d-1)), corresponding to a single user request. Instead of visualizing the entire execution as one continuous timeline, \name further decomposes each task into a sequence of trace cards, each representing a subtask or sub-agent execution. This organization preserves workflow context while preventing long-running executions from becoming difficult to inspect.

\begin{figure*}[t]
\centering
\includegraphics[width=\linewidth]{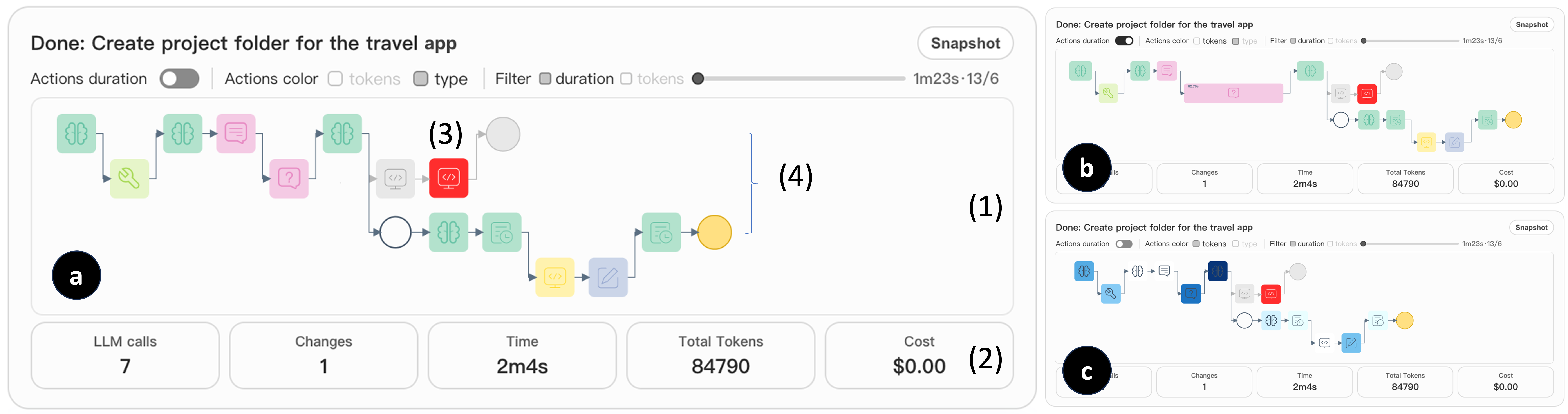}
\caption{Visualization of an execution shown in a trace card.}
\label{fig:visual}
\vspace{-1.0em}
\end{figure*}

\textbf{Trace Cards.}
Trace cards are the fundamental representation unit in \name (Fig.~\ref{fig:visual}). Each card summarizes one execution stage through two complementary components: an \textit{Execution Flow (Fig.~\ref{fig:visual}(a-1))} that explains how the agent performs the task and an \textit{Execution Summary (Fig.~\ref{fig:visual}(a-2))} that characterizes the corresponding execution cost. Together, they enable users to understand both the execution process and its efficiency:

\underline{Execution Flow.}
The execution flow visualizes the temporal trajectory of agent behaviors as a directed workflow beginning with the user request and ending with the execution outcome. Intermediate nodes represent primitive agent actions extracted in Section~\ref{sec:extraction}, while directed edges encode temporal and causal dependencies between actions.
Each node visually encodes key execution attributes, with semantic icons representing action types and visual variables (i.e., node size and opacity) indicating execution duration (Fig.~\ref{fig:visual}(b)) and token consumption (Fig.~\ref{fig:visual}(c)), respectively. Failed actions are highlighted in red (Fig.~\ref{fig:visual}(a-3)) to facilitate rapid identification of execution failures.
To show execution revisions, \name represents redirected executions as alternative branches that diverge from the selected action while preserving the shared execution history (Fig.~\ref{fig:visual}(a-4)). By presenting the original and redirected trajectories within the same execution graph, users can directly compare their behavioral differences, understand how interventions propagate through subsequent execution steps, and evaluate alternative execution outcomes without losing previously validated execution context.

\underline{Execution Summary.}
Complementing the execution flow, each trace card presents aggregated execution statistics, including runtime, token consumption, monetary cost, and the number of LLM invocations. These summaries provide an at-a-glance overview of execution efficiency, enabling users to rapidly identify computationally expensive or abnormal executions before inspecting detailed traces.

\subsubsection{Trace Interaction}
Beyond representing execution, \name transforms execution traces into interactive control surfaces that support both execution inspection and trajectory-level intervention, enabling users to progressively steer agent behaviors (DG3).

\textbf{Inspection.}
Inspection supports a progressive exploration of agent execution through four complementary interaction techniques, enabling users to move from observing overall execution to inspecting individual actions:
(i)~\textit{Streaming.} Execution traces are streamed incrementally as the agent executes, allowing users to monitor execution progress in real time and promptly detect unexpected behaviors.
(ii)~\textit{Cross-view Linking.} After identifying an execution stage of interest, users can select a trace card or an action node to automatically link to the corresponding conversation and execution records in the Dialog Panel (Fig.~\ref{fig:ui}(d-3)), enabling seamless navigation between high-level workflows and low-level execution details.
(iii)~\textit{Cost Filtering.} To further narrow the inspection scope, users can filter execution nodes by runtime or token consumption, allowing computationally expensive or abnormal execution steps to be quickly located in long-running tasks.
(iv)~\textit{Contextual Inspection.} Finally, hovering over a trace element reveals contextual information, including action parameters, execution outputs, and resulting states, enabling detailed inspection without disrupting the overall workflow visualization.
Together, these interaction techniques support a coarse-to-fine exploration of execution traces, helping users efficiently develop a causal understanding of agent behaviors.

\textbf{Intervention.}
Building upon inspection, \name enables users to progressively steer agent execution through three complementary intervention techniques spanning different levels of granularity.
(i)~\textit{Comment.}
Users can annotate individual execution traces with preferences, observations, or issues, providing localized feedback that is subsequently incorporated into trace-to-skill distillation.
(ii)~\textit{Redirect.}
When localized feedback is insufficient, users can fork execution from any previous action node and continue along a revised trajectory. By reusing all validated execution before the fork point, Redirect enables efficient correction without restarting the entire workflow.
(iii)~\textit{Outcome Correction.}
After task completion, users can provide feedback on the final execution outcome, complementing step- and trajectory-level interventions with task-level supervision.
Together, these intervention techniques transform execution traces from passive records into editable execution trajectories, enabling users to iteratively refine agent behaviors while preserving previously validated progress.

The resulting interaction traces, together with users' explicit interventions, serve as the foundation for the trace-to-skill distillation mechanism described in the next section.

\begin{figure*}[t]
\centering
\includegraphics[width=\linewidth]{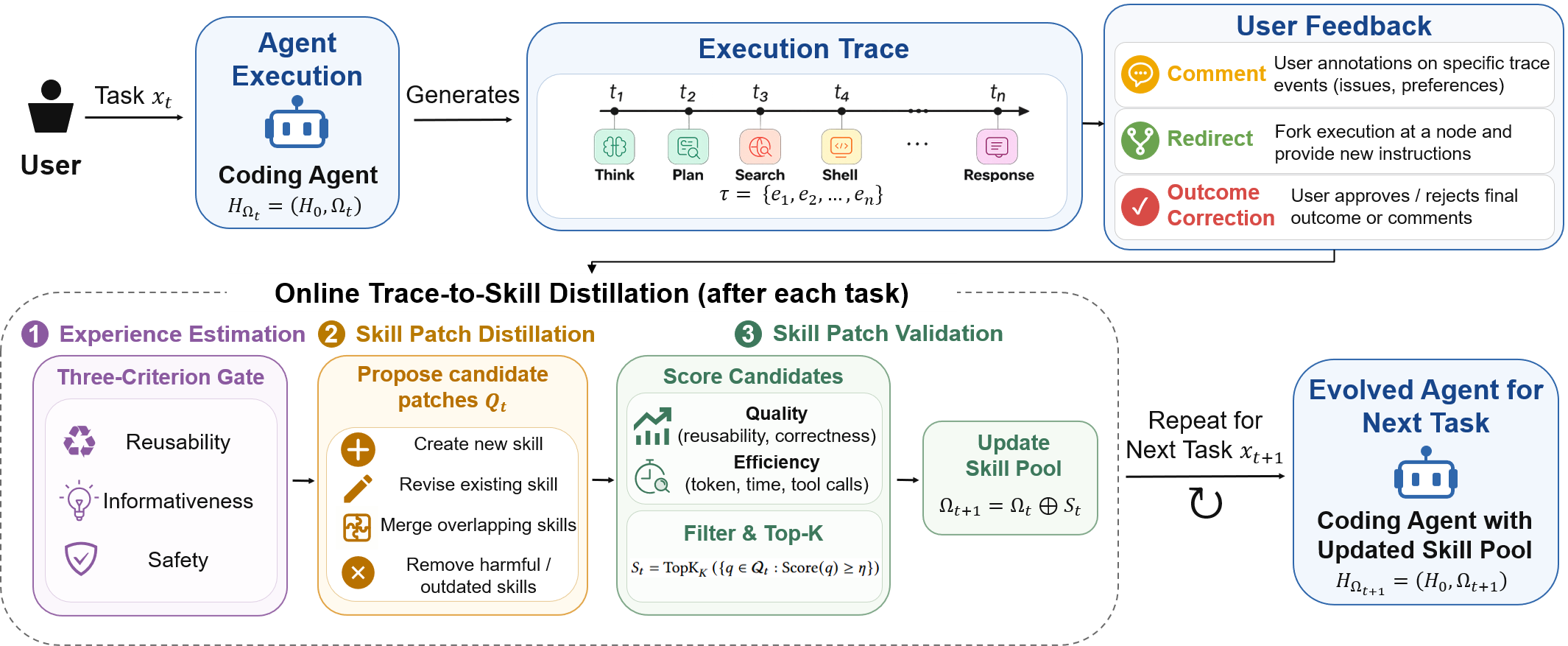}
\caption{Overview of the online trace-to-skill distillation process in \name. 
Execution traces and human feedback are analyzed to extract reusable skill patches, 
which are evaluated and integrated into the skill pool to continuously improve future agent behaviors.}
\label{fig:algorithm}
\end{figure*}

\subsection{\bf Online Trace-to-Skill Distillation}
\label{sec:Trace-to-Skill}
Trace-based visualization and intervention enable users to actively steer coding agents during individual tasks. However, such collaboration knowledge is often transient: successful corrections, debugging strategies, and workflow preferences disappear once a session ends. Consequently, the agent repeatedly makes similar mistakes even when users have already demonstrated better solutions.
\name addresses this limitation by transforming human interventions into reusable skills that continuously improve future executions. The key insight is that user feedback should not be memorized as execution trajectories. Instead, it provides evidence about \emph{how the agent should behave under similar situations}. For example, when a user redirects an execution to inspect the repository before modifying code, the reusable knowledge is not the corrected trajectory itself, but the behavioral strategy of inspecting existing implementations before editing.

Based on this insight, \name models skill evolution as a human-guided knowledge refinement process. As illustrated in Fig.~\ref{fig:algorithm}, each interaction experience goes through three sequential stages:
\textbf{\textit{Experience Estimation}} determines whether an interaction contains reusable knowledge.
\textbf{\textit{Skill Patch Distillation}} extracts transferable behavioral changes from accepted interactions.
\textbf{\textit{Skill Patch Validation}} evaluates candidate patches and incorporates only high-quality ones into the evolving skill pool.
Together, these stages continuously transform transient human interventions into persistent collaboration knowledge without modifying the underlying foundation model.

\textbf{Evolution Objective.} We model the coding agent as
\begin{equation}
H_{\Omega}=(H_0,\Omega),
\end{equation}
where $H_0$ denotes the fixed foundation model together with its tools and execution environment, while $\Omega$ represents the evolving skill pool.
Given a task stream $\{x_1,x_2,\ldots\}$, after completing task $x_t$, the agent updates the skill pool using the execution trace $\tau_t$ and the associated user feedback $f_t$:
\begin{equation}
\Omega_{t+1}
=
\operatorname{Update}
(\Omega_t,\tau_t,f_t).
\end{equation}

Unlike model fine-tuning, \name never modifies the parameters of $H_0$. Instead, it achieves continual adaptation by incrementally refining the external skill pool.
The objective of evolution is to improve future task quality while reducing execution costs. Let
$\mathcal{M}=\{\mathrm{token},\mathrm{time},\mathrm{call}\}$ denote the three execution-cost dimensions.
The overall utility of a skill pool is defined as
\begin{equation}
\label{eq:skill-objective}
J(\Omega)
=
\mathbb{E}_{x\sim\mathcal{X}}
\left[
Q(\tau_x^\Omega)
\prod_{m\in\mathcal{M}}
(1-\bar c_m(\tau_x^\Omega))
\right],
\end{equation}
where 
$\tau_x^{\Omega}$ is the execution trace produced by the agent based on skill pool $\Omega$ on task $x$; $Q(\tau_x^\Omega)\in[0,1]$ measures task quality using user feedback and available environment verification, and $\bar c_m(\tau_x^\Omega)\in[0,1]$ denotes the normalized execution cost along dimension $m$.
Intuitively, Equation~(\ref{eq:skill-objective}) rewards skill pools that simultaneously improve task quality while reducing token consumption, execution time, and tool usage. Rather than optimizing this objective directly, \name approximates it through the three-stage evolution pipeline described below.

\textbf{Experience Estimation.}
Not every user intervention should become a reusable skill. Some interactions are highly task-specific, ambiguous, or even misleading. Learning from such experiences may negatively affect future executions.
Therefore, before generating any skill, \name first estimates whether an interaction contains reusable evolution signals. Specifically, each trace-feedback pair is evaluated from three complementary perspectives: (i)~\textit{Reusability:} can the observed behavior generalize beyond the current task?
(ii)~\textit{Informativeness:} does the feedback imply a concrete behavioral improvement?
(iii)~\textit{Safety:} can the behavior be reused without violating user constraints or introducing undesirable side effects? To this end, \name introduces an LLM-based estimator $\operatorname{Estimate}(\cdot)$ to predict the reusability, informativeness, and safety scores from the trace $\tau_t$ and user feedback $f_t$:
\begin{equation}
\mathbf{z}_t
=
\operatorname{Estimate}
(\tau_t,f_t)
\in[0,1]^3,
\end{equation}

\textbf{Skill Patch Distillation.}
After an intervention passes the estimation stage, \name distills the observed behavior into reusable knowledge.
Instead of memorizing the corrected execution trace, the distillation process extracts the behavioral difference between the original and corrected executions. A reusable skill should answer three questions:
(i)~\textit{When} should the behavior be activated (\emph{Context})?
(ii)~\textit{What} behavior should change (\emph{Decision})?
(iii)~\textit{How} should future execution proceed (\emph{Strategy})?
For example, when a user redirects the agent to inspect the repository before modifying code, the distilled skill records the reusable strategy rather than the corrected trajectory itself.
The distillation process generates a set of candidate skill patches:
\begin{equation}
\mathcal{Q}_t
=
\operatorname{Distill}
(\tau_t,f_t,\Omega_t),
\end{equation}
where $\operatorname{Distill}(\cdot)$ is implemented using a large language model. Rather than producing standalone skills, each candidate
\(
q=(\mathrm{op},\psi,\pi)\in\mathcal{Q}_t
\)
represents a \emph{skill patch}---a structured modification to the current skill pool. Here, $\mathrm{op}$ is one of four update operations: \textbf{\textit{Create}} (introduce a new skill), \textbf{\textit{Revise}} (update an existing skill), \textbf{\textit{Merge}} (consolidate overlapping skills), or \textbf{\textit{Remove}} (discard obsolete or conflicting knowledge); $\psi$ encodes the distilled behavioral knowledge (context, decision, and strategy); and $\pi$ records the originating execution trace and user feedback for provenance.



\textbf{Skill Patch Validation and Skill Pool Evolution.}
Although estimator-approved interactions provide reliable evidence, the resulting skill patches may still be overly specialized, redundant, or offer limited practical benefit.
Before updating the skill pool, \name uses an LLM-based scorer to evaluate each patch through \textit{quality assessment} followed by \textit{efficiency estimation}.

For a patch $q$, the scorer first predicts
$\mathbf{h}(q)\in[0,1]^2$, representing its expected \textit{reusability} and \textit{correctness}. Only patches satisfying the quality thresholds $\boldsymbol{\alpha}$ are retained. For each qualified patch, the scorer further predicts its expected efficiency improvement $\boldsymbol{\rho}(q)\in[-1,1]^3$ over token consumption, execution time, and tool calls, where positive values indicate expected cost reduction and negative values indicate additional overhead.
Quality and efficiency are jointly combined into the patch score:
\begin{equation}
\operatorname{Score}(q)
=
\mathbb{I}
(\mathbf{h}(q)\succeq\boldsymbol{\alpha})
\left(
\prod_{m\in\mathcal{M}}
\max(0,\rho_m(q))
\right)^{1/3},
\end{equation}
where the quality gate ensures that efficiency gains cannot compensate for unreliable knowledge, while the geometric mean prioritizes patches that consistently reduce execution costs across all three dimensions.
Finally, patches are ranked by their scores, and the highest-scoring updates are selected to modify the skill pool:
\begin{equation}
\mathcal{S}_t
=
\operatorname{TopK}_K
\left\{
q\in\mathcal{Q}_t
\mid
\operatorname{Score}(q)\ge\eta
\right\}.
\end{equation}
\begin{equation}
\Omega_{t+1}
=
\Omega_t
\oplus
\mathcal{S}_t,
\end{equation}
where $\oplus$ denotes applying the selected create, revise, merge, and remove operations. If $\mathcal{S}_t=\varnothing$, the skill pool remains unchanged.
Rather than continuously accumulating new skills, \name maintains the skill pool through iterative refinement. By repeatedly creating, revising, merging, and removing skills, transient human interventions gradually become persistent collaboration knowledge that guides future executions. Consequently, the coding agent evolves through an increasingly capable human-guided skill pool instead of modifying the parameters of its foundation model.

\section{Evaluation}

We first provide a quantitative evaluation of the online trace-to-skill algorithm outlined in Section~\ref{sec:Trace-to-Skill} in an open-source coding benchmark, and then a user study for evaluating the effectiveness of the visualization design.
Last, we present a user study with eight experts to evaluate \name in real-world coding tasks and demonstrate usage cases.

\subsection{Skill-Based Agent Evolving Mechanism}

We conduct a quantitative evaluation of the proposed online trace-to-skill distillation mechanism to highlight the \name's ability of evolving coding agents through trace-based interaction (DG4).
Specifically, we ask whether skills distilled online from prior tasks improve accuracy and efficiency on subsequent related tasks, and whether online distillation from user feedback provides additional benefit over learning from execution traces alone.

\textbf{Dataset and Metrics.}
We use SWE-bench Verified~\cite{jimenez2024swebench}, a human-validated benchmark of real-world GitHub issues with executable test suites. 
To study skill transfer across related tasks, we cluster tasks by repository and  subsystems, inferred from the files modified by the reference patch, and split each cluster into distillation and held-out test tasks. We then sample tasks from each retained difficulty, prioritizing distillation tasks from clusters represented in the test set. The result data contains 30 distillation tasks and 20 held-out tasks. 
Skill evolution starts from an empty pool, updates it sequentially over the distillation stream, and then freezes the learned pool for evaluation on the held-out set.

We consider the metrics regarding both task completion and resource consumption of AI, including:
\begin{itemize}[leftmargin=0.5cm]
    \item \textit{Pass rate} : The proportion of tasks accepted by the official SWE-bench verifier.
\item \textit{Input tokens}: The total number of cached input tokens, read tokens, and write tokens.
\item \textit{Output tokens}: This includes both generated response tokens and reasoning tokens.
\item \textit{Cost}: This metric translates the different token categories into actual financial cost.
\item \textit{Time}: The duration taken to execute a task.
\item \textit{The overall utility $J(\Omega)$} (Equation~\ref{eq:skill-objective}): Its quality term assigns unit quality to a verifier-accepted trace and zero quality otherwise, while its token-consumption term uses \textit{Cost} to account for the different prices of uncached input, cached input, and output tokens.
\item \textit{Tool calls}: The count of tool invocations during the task.
\end{itemize}

\textbf{Baselines.}
We compare our method with two baselines using the same coding agent and evaluation harness.
\textbf{Base} has no distilled skills, so the agent has no persistent experience for new tasks.
\textbf{Trace-only} evolves skills from execution traces without user feedback, updating the skill pool after each task with the same skill-distillation framework as our method.
Our method, \textbf{Trace-fb}, simulates two forms of feedback using the benchmark verifier.
For \textit{Redirect}, when the verifier marks the original execution as unresolved, we fork the trace with the user hint \textit{Locate the root cause more carefully and make a minimal, correct fix; avoid unrelated, broad changes.}''. If the verifier marks the fork as resolved, the abandoned and corrected branches form a pairwise feedback. For \textit{Outcome Correction}, we append \textit{I've reviewed the result; the requested behavior is resolved.}'' to a trace accepted by the verifier and the correction above to a rejected trace.
We do not additionally involve \textit{Comment} as it follows the same processing as \textit{Outcome Correction}.

\textbf{Implementation.}
We use DeepSeek-V4-Flash~\cite{deepseekai2026deepseekv4} as the shared LLM backbone for task execution and all three skill-distillation stages.
Each task starts from a clean repository checkout with isolated OpenCode state.
External skills are disabled, and verification uses the official Docker-based SWE-bench harness.
The estimator, distiller, and scorer are each executed in a separate one-shot OpenCode session, isolated from the task-execution session and from one another, to prevent conversational state from leaking across modules or tasks. 
The estimator thresholds for reusability, informativeness, and safety are 0.34, 0.50, and 0.60. Candidate patches must reach 0.50 in both reusability and correctness, predict positive savings in all three resource dimensions, and obtain an aggregate score of at least 0.10. At most the top three patches are applied per task. The cost/time/tool-call budgets are $(0.02,600,60)$, $(0.04,900,90)$, and $(0.05,1200,100)$ for the three retained difficulty levels. Trace-only and Trace-fb evolve separate skill pools under the same settings.

\begin{table*}[t]
\centering
\caption{Held-out SWE-bench Verified results over 20 tasks and five repeated runs. The split and stream order use seed 42. Values are mean $\pm$ population standard deviation. Tokens are in thousands and cost is in $10^{-3}$~\$. Parentheses in the input-token column show mean cache-read tokens.}
\label{tab:skill-evolution-results}
\scriptsize
\setlength{\tabcolsep}{3pt}
\resizebox{\textwidth}{!}{%
\begin{tabular}{lccccccc}
\toprule
Method & Pass (\%) $\uparrow$ & In-Tok. (Cached) $\downarrow$ & Out-Tok. $\downarrow$ & Cost ($10^{-3}$~\$) $\downarrow$ & Time (s) $\downarrow$ & Calls $\downarrow$ & $J(\Omega)$ $\uparrow$ \\
\midrule
Base
& $65.00\!\pm\!3.16$ & $1459.60\!\pm\!467.94$ ($1431.45$) & $17.62\!\pm\!0.97$ & $12.88\!\pm\!1.89$ & $268.25\!\pm\!15.76$ & $38.26\!\pm\!3.91$ & $0.25\!\pm\!0.01$ \\
Trace-only
& $65.00\!\pm\!4.47$ & $1303.26\!\pm\!284.74$ ($1275.60$) & $16.43\!\pm\!2.40$ & $12.04\!\pm\!1.66$ & $257.05\!\pm\!25.05$ & $36.35\!\pm\!3.83$ & $0.26\!\pm\!0.02$ \\
\textbf{Trace-fb (ours)}
& $\mathbf{68.00\!\pm\!2.45}$ & $\mathbf{1123.33\!\pm\!137.32}$ ($\mathbf{1098.68}$) & $\mathbf{15.56\!\pm\!1.34}$ & $\mathbf{10.88\!\pm\!0.94}$ & $\mathbf{229.32\!\pm\!16.00}$ & $\mathbf{34.47\!\pm\!2.64}$ & $\mathbf{0.28\!\pm\!0.04}$ \\
\bottomrule
\vspace{-3em}
\end{tabular}%
}
\end{table*}

\textbf{Results.}
Table~\ref{tab:skill-evolution-results} shows that our method performs best in all seven metrics.
It improves both quality and efficiency.
Compared to \textbf{Base}, our approach increases the pass rate from 65.0\% to 68.0\%, reduces input tokens by 23.0\%, and lowers cost and time by 15.5\% and 14.5\%.These improvements result in a 14.0\% enhancement in $J(\Omega)$.
Furthermore, our method consistently surpasses \textbf{Trace-only}, which only focuses on improving efficiency.
This outcome indicates that feedback not only enhances quality by utilizing human corrections as signals but also supports more targeted skill-pool updates for better efficiency.

\subsection{The Trace Visualization Design}

To evaluate whether \name helps users better understand agent execution processes and locate errors, we conducted a controlled study comparing trace visualizations from \name with OpenCode's native logs.

\textbf{Participants.}
We recruited $20$ participants with coding-agent experience (age: M=$24.1$, SD=$3.7$; $13$ females). Participants were randomly assigned to two groups, each completing one of the two questionnaires.

\textbf{Materials.}
We collected 10 agent execution traces from real coding sessions. Each trace captures a portion of the agent's dialogue history and execution steps. Based on each trace, we designed one multiple-choice question targeting execution process comprehension, error localization, or root cause analysis. Each trace was presented in two formats: (i)~\textit{Baseline}: Direct output from OpenCode's native terminal user interface, and (ii)~\name's trace visualization. 
This led to 10 questions with agent execution traces paired with 2 presentation formats each. 
We created two counterbalanced questionnaire versions, each containing all 10 questions with five per format. 

\textbf{Procedure.}
The experiment was conducted through an online questionnaire platform. Participants first read a brief introduction to the \name and OpenCode user interfaces. They then completed 10 multiple-choice questions in sequence. The system automatically recorded answers and response time for each question.

\textbf{Results.}
\begin{figure*}[t]
  \centering
  \includegraphics[width=\textwidth]{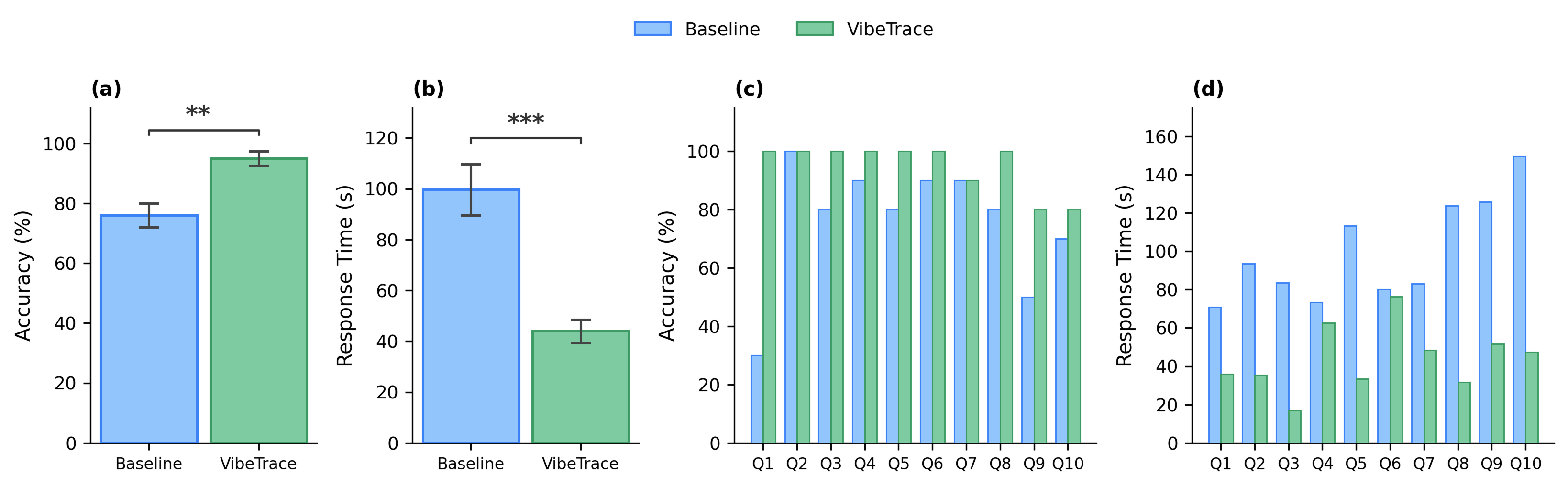}
  \caption{Accuracy and response time comparison between \name and the baseline; **~denotes $p<0.01$ and ***~denotes $p<0.001$. (a)~Overall accuracy: \name\ (95\%) and the baseline (76\%). (b)~Average response time: \name\ (44.0\,s) and the baseline (99.7\,s). (c)~Per-question accuracy across Q1--Q10. (d)~Per-question response time across Q1--Q10.}
\label{fig:questionnaire}
\end{figure*}
We analyze the study results based on the accuracy and the response time per question (Fig.~\ref{fig:questionnaire}). \name\ (M=95.0\%, SD=11.4) achieves significantly higher accuracy than the baseline (M=76.0\%, SD=17.9), as confirmed by the Wilcoxon signed-rank test~\cite{woolson2007wilcoxon} ($|Z|=3.09$, $p<0.01$). In terms of response time (Fig.~\ref{fig:questionnaire}b), \name\ (M=44.0\,s, SD=20.6) is also significantly faster than the baseline (M=99.7\,s, SD=45.2) ($|Z|=3.92$, $p<0.001$). These results suggest that the trace visualization proposed by \name\ can more effectively and efficiently support users in understanding agent execution traces, localizing errors, and analyzing root causes.

\subsection{System Usability}
\label{sec:userstudy}

To evaluate how \name supports human--coding agent collaboration, we conducted a third experiment comprising case studies and interviews with expert users. We examined four aspects of collaboration: whether users could (i) observe the agent's ongoing and completed activities efficiently (\textbf{DG1}); (ii) understand the agent's decisions and identify potential errors (\textbf{DG2}); (iii) steer the agent effectively based on this understanding (\textbf{DG3}); and (iv) evolve the agent by distilling reusable knowledge from their interactions (\textbf{DG4}).

\textbf{Participants.}
We recruited eight expert users from diverse backgrounds, including both researchers and industry practitioners (age: M=25.5, SD=4.1; four females). All participants had expertise and solid experience in using coding agents such as Cursor, Claude Code, or Copilot, 3 had built agents of their own, providing a deeper understanding of agent execution. This composition allowed us to observe \name across users with varying levels of coding and agent expertise. 


\textbf{Procedure.}
The study consists of three phases: onboarding, tasks, and an interview. Participants were asked to share their screen during the study. Each session was recorded and transcribed for subsequent analysis, and informed consent was obtained from all participants.

\textit{Onboarding.}
We began with a brief overview of the study procedure and assisted participants in setting up \name on their own machines. 
Then, we played an 8-minute demonstration video that highlighted \name's core features to help them get familiar with the system. We also collected the pre-study demographic questionnaire. 

\textit{Tasks.} Participants were requested to use \name to complete two vibe-coding tasks. The two tasks were designed to be complementary. 
The first is a lightweight data analysis task that aims to uncover the return-on-investment patterns of the film industry over the past two decades, finding which genres achieve the highest ROI and whether these profitable genres also hold an advantage in audience ratings, and finally presenting the results as an HTML data report. 
The second is a debugging-heavy task that depends on multiple external services and aims to build a web application that converts an uploaded image into a bead-art pattern annotated with color codes. In both tasks, participants were asked to interpret what the agent was doing and how it proceeded, to judge whether errors arose and, if so, analyze their causes and decide whether to intervene, and ultimately to complete the task through several rounds of iteration. 
Participants were asked to \emph{think aloud} throughout and were free to ask questions at any time. In parallel, \name logged fine-grained behavioral traces, including the timing of panel openings and the sequence and frequency of feature invocations, providing a quantitative complement to the think-aloud data for characterizing participants' usage patterns.
After the two assigned tasks, we invited participants to continue using \name in their own day-to-day work and to share their experiences with us. We collected the resulting real-world usage traces.

\textit{Interview.} At the end of the tasks, participants completed the System Usability Scale (SUS)~\cite{brooke1996sus} and took part in a 15-minute open-ended discussion about \name's observability, explainability, controllability in the human---coding agent collaboration. They were also asked with their suggestions for improvement. 

\textbf{Results.} We report three sets of results: the System Usability Scale scores, the logged usage patterns across the four panels, and the interview findings.
\begin{figure}[t]
  \centering
  \includegraphics[width=\columnwidth]{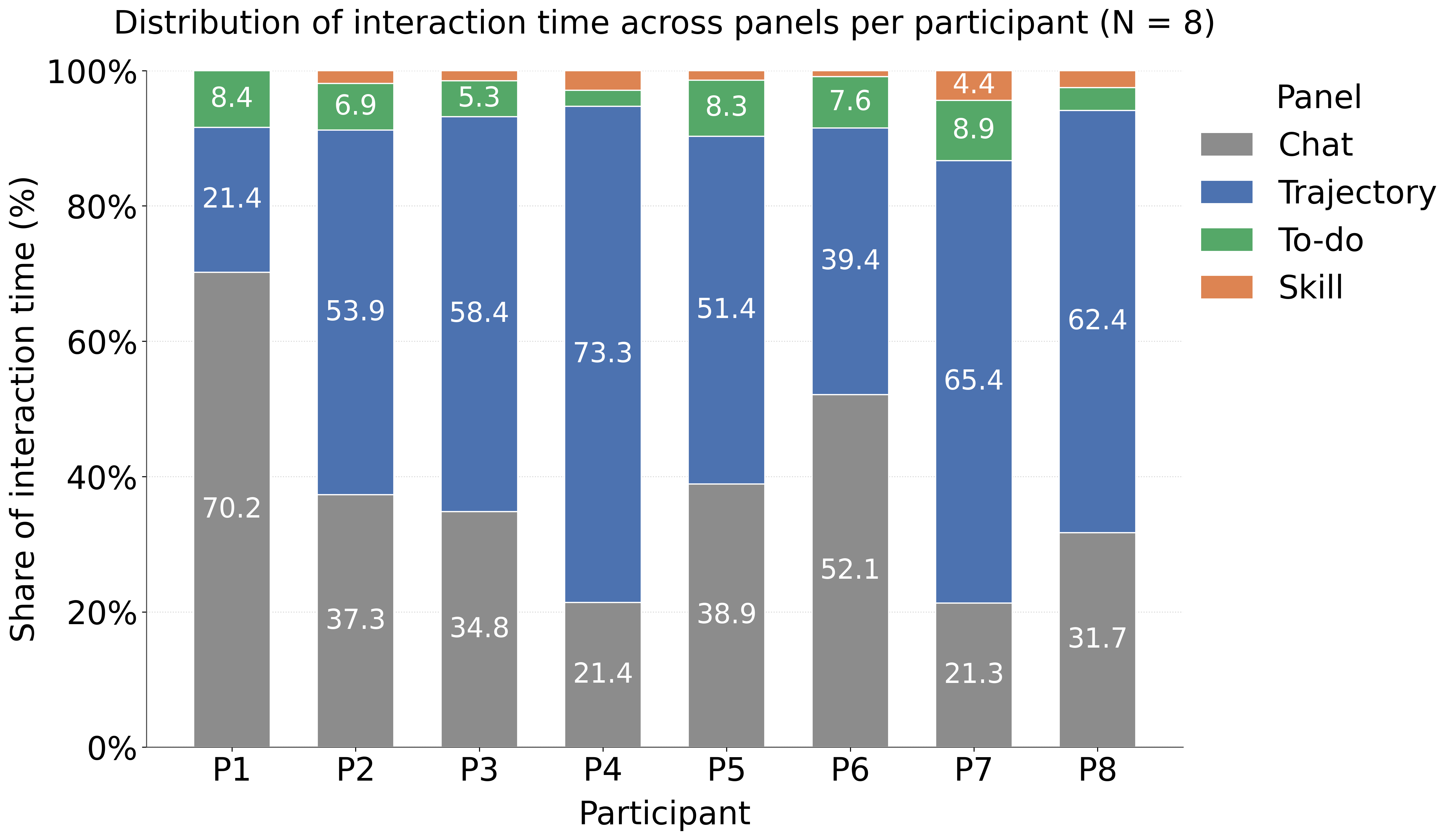}
  \caption{Distribution of interaction time across the four panels for each participant ($N=8$).}
  \label{fig:study-results}
\end{figure}
\begin{figure*}[t]
  \includegraphics[width=\textwidth]{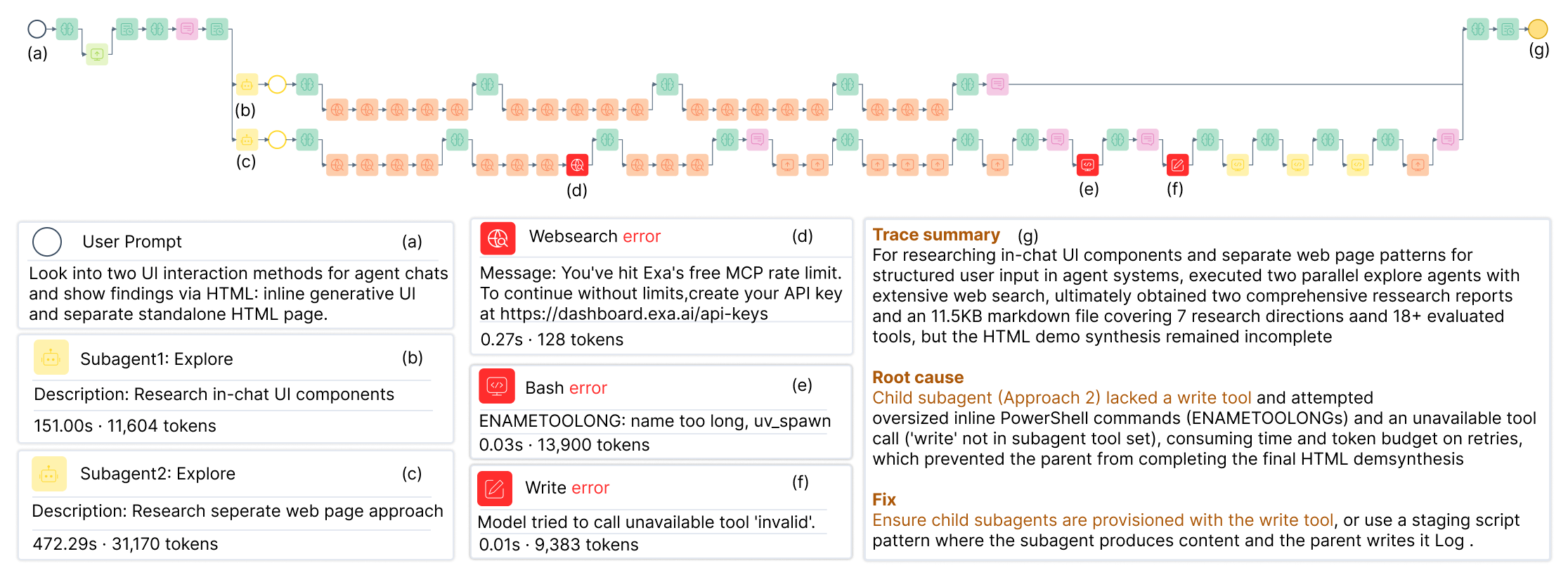}
  \caption{\name supports observing and understanding long-running agent execution through interactive visualization of agent execution traces. The figure illustrates a representative workflow: (a) the user asks the agent to investigate two UI techniques and synthesize the findings into a webpage; (b--c) the interaction trace branches into two subagent workflows, each exploring one technique; (d--f) the visualization reveals three failures in the second branch, including a rate-limit error, an oversized argument list, and an invalid tool invocation; (g) the terminal state summarizes the execution, identifies the root causes, and recommends corresponding fixes.}
  \label{fig:case1}
\end{figure*}

\textit{System Usability Scale (SUS).}
\name achieved a mean SUS score of $M=85.9$ ($SD=9.4$), indicating strong functional integration, interaction consistency, and ease of operation. Decomposed under the two-factor structure of Lewis and Sauro~\cite{lewis2009factor}, \name scores $M=89.4$ on the eight-item Usability subscale and $M=71.9$ on the two-item Learnability subscale. Several participants reported that setup, not the interface, was the main barrier to first use. A subsequent release addresses this: \name now installs as a single-command OpenCode plugin that launches its services automatically.

\textit{Usage Patterns.}
We analyzed the usage patterns of participants by measuring the proportion of interaction time spent on each panel of \name (Fig.~\ref{fig:study-results}). The numbers were measured by cursor movement. 
Most participants were highly engaged with the Trace panel. Six participants dedicated the majority of their time to it. This indicates that the visualization was effectively utilized during their interaction. Two participants (P1 and P6) spent more than $50\%$ of their time on the Dialog panel, with the Trace panel being their second most-used feature. 
Overall, participants agreed that the trajectory visualization was the most helpful component of \name.

\textit{Interviews.}
Participants provided positive feedback regarding \name's observability, explainability, and controllability in human--agent coding collaboration.
When observing a long and complex workflow, participants appreciated the visualization provided in the Trace panel. For instance, P6 said, ``The tree structure clearly lays out the task flow and the workflow, and the linkage between panels lets me quickly locate the step I want to view.''
When an execution encountered an error, participants found it convenient to understand with \name. P5 stated, ``When something goes wrong, I can easily see the content before and after the error point, which makes locating and analyzing it convenient.'' P3 said, ``The output of the middle nodes helps me understand.''
Participants valued the ability to intervene in the agent's execution. P7 stated, ``Ask about the current state more clearly, without having to stop it.'' P8 noted, ``Quickly see which steps are error-prone, and find a relevant skill to remedy those errors.''
Additionally, participants appreciated the online skill distillation mechanism. P4 mentioned, ``A skill summary is really necessary; this knowledge used to come from human instructions, but I couldn't follow the trial and error inside the command line, so having it distilled automatically is very helpful.'' P3 said, ``There's no need to explain some processes over and over.''

\subsection{Example Cases}
\label{sec:cases}
We present three example cases from participants' own projects in the expert user study (Section~\ref{sec:userstudy}) to illustrate how \name supports observing, understanding, intervening in, and improving human--agent collaboration.

\subsubsection{Observing and Understanding Agent Execution.}
This case demonstrated how P7 uses \name to do technical investigation. The task is to compare two candidate user interface techniques for an agent chat application: one embeds UI components directly in the chat, and one renders the selection on a separate web page and returns the result. 
The user wants to synthesize the comparison into an HTML webpage.
This task is long-running and requires multiple steps of reasoning and tool calls (see Fig.~\ref{fig:case1}).
After the user prompt (Fig.~\ref{fig:case1}-a), the agent automatically explores the two techniques in parallel, one for in-chat UI components (Fig.~\ref{fig:case1}-b) and one for the separate web page (Fig.~\ref{fig:case1}-c), where the overall structure is clearly visualized in the trace visualization panel (DG1).
When the user checks the execution of the two branches, he identifies three errors highlighted in red in the second branch, which might affect the final result. Therefore, the user clicks each error node to view the details one by one.
He reads that the first error is a rate limit from a web search tool (Fig.~\ref{fig:case1}-d), but the following web search actions complete normally, and no downstream effect propagates to the overall execution.
Then, the user reads the remaining two errors: the second error is an oversized argument list from a shell command (Fig.~\ref{fig:case1}-e), and the third error is an invalid tool call (Fig.~\ref{fig:case1}-f). 
Rather than traversing each corresponding node, the user clicks the final node to check the summary of the execution (Fig.~\ref{fig:case1}-g). He understands that the agent has completed the technical investigation but failed to synthesize the results into a webpage due to the previous errors.
The root cause is that the agent was not provisioned with the \texttt{write} tool, so it first attempted to persist its markdown notes by inlining the entire body into a single \texttt{shell} command, whose over-long argument list was rejected. Then, a subsequent call to the \texttt{write} tool failed for the same reason (DG2).
The way to fix the errors is to grant the agent \texttt{write} access, so that the tasks can save their outputs directly.
Through \name, the user can easily observe the agent execution trace (\textbf{DG1}), and understand its execution and failures during a long-running task (\textbf{DG2}).

\begin{figure*}[t] \centering \includegraphics[width=\linewidth]{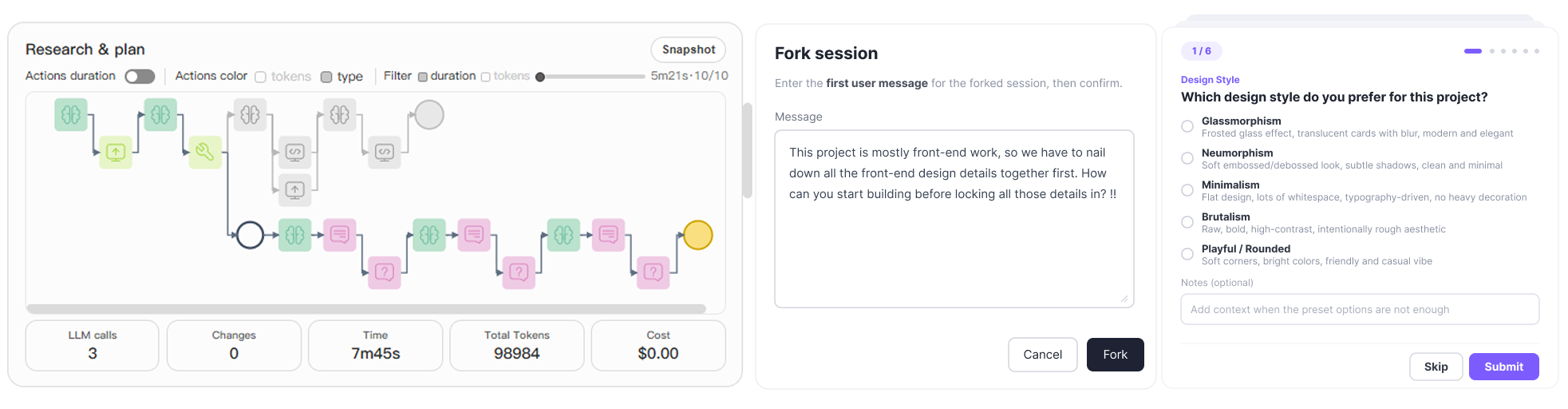} \caption{\name lets the user intervene in the agent execution by forking the execution trace. Left: The user forks a new branch before development begins, prompting a request for design clarification. The greyed-out branch is the original trace that implements the application without consulting design styles, while the lower branch is the newly forked branch that includes user feedback. Middle: the user feedback interface of the forked branch for reminding the agent to settle the frontend design details before building. Right: On the forked branch, the agent uses the \texttt{question} tool to ask design details.} 
\label{fig:case2} 
\end{figure*}

\subsubsection{Intervening Agent Execution.}
This case demonstrated a key intervening process when P2 builds an application via \name. It's a simple web application generating meal suggestions.
The user prompts the agent to build a frontend demo, and the agent scaffolds the entire project step by step and delivers the final implementation.
However, the user finds that the agent has chosen the wrong visual style and color palette without consulting the user.
Rather than terminating the session and re-prompting from scratch, the user forks a new branch immediately before the node where the development begins (Fig.~\ref{fig:case2}, left). 
The user deliberately selects that node: the earlier discussion that fixed the feature set is the valuable groundwork he wants to carry forward rather than reconstruct, while the design decisions the agent has since committed to are exactly what he wants to change.
On the forked branch, the user prompts the agent to align on the design styles before implementing anything (Fig.~\ref{fig:case2}, middle). 
Given that instruction, the agent then invokes the \texttt{question} tool several times to request design clarification from the user (Fig.~\ref{fig:case2}, right). The resulting implementation in this forked branch better satisfies the user's preferences.
Through \name, the user is able to precisely intervene in the agent execution trace to align the agent's behavior with user expectations, thus improving the efficiency of the collaboration (\textbf{DG3}).

\begin{figure*}[t]
\centering
\includegraphics[width=\linewidth]{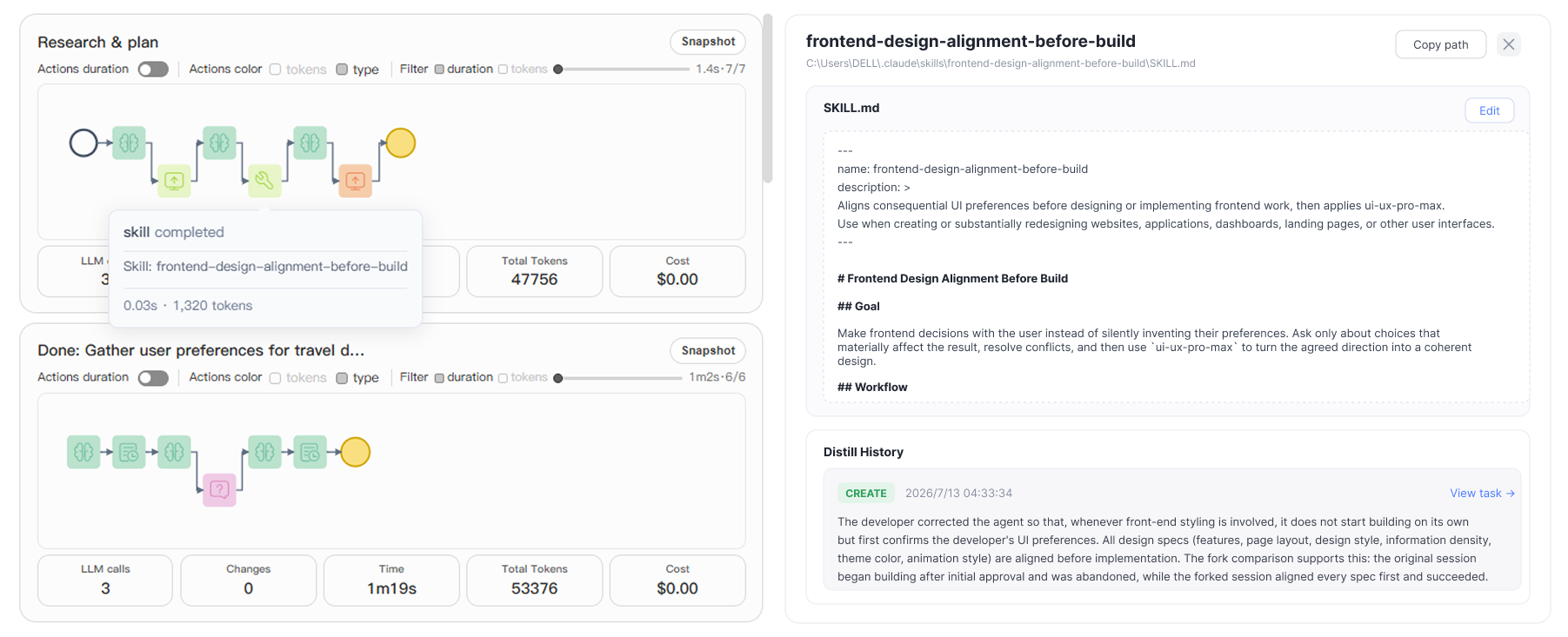}
\caption{\name distills a skill from human feedback and reuses it in a later task. Left: the agent automatically matches the distilled skill. Right: \name automatically distilled \texttt{front-end-design-alignment-before-build.skill}.}
\label{fig:case3}
\end{figure*}

\subsubsection{Agent Self-Evolving from Human Feedback.}
After the user forks the trace and accept the changes in Case 2, \name automatically distills a new \texttt{front-end-design-alignment-before-build} skill by comparing the traces before and after the fork. 
According to the skill description (Fig.~\ref{fig:case3}, right top), whenever a task involves frontend development and the visual direction is not yet fixed, the agent should use the \texttt{question} tool to confirm design styles with the user before any implementations.
The Distill History (Fig.~\ref{fig:case3}, right bottom) describes how that skill was distilled from human feedback.
When the user works with \name on another similar frontend project, the agent automatically matches the skill to confirm design styles with the user before building anything (Fig.~\ref{fig:case3}, left).
This demonstrate how \name learn from human intervention and distill a reusable skill that can be applied to future tasks (\textbf{DG4}).

\section{Discussion}

Our evaluation demonstrates that \name can improve both the efficiency of agent execution and users' ability to observe, understand, and steer coding agents. Beyond these results, the studies reveal several design tensions in human--agent collaboration and suggest broader principles for designing agentic coding systems.

\subsection{Design Tensions Revealed by User Feedback.}
Rather than treating all user behaviors as design failures, our interviews reveal several natural tensions that arise when users collaborate with increasingly autonomous coding agents.

\textbf{Selective oversight vs. comprehensive understanding}
Participants did not continuously monitor the entire execution trace. Instead, they typically started from the Dialog Panel, used the task overview to locate a relevant trajectory segment, and then inspected details in the Trace Panel. This pattern suggests that users often seek answers to specific questions rather than reconstructing the complete execution history. As P6 noted, ``the tree structure clearly lays out the task flow and the workflow, and the linkage between panels lets me quickly locate the step I want to view.'' Thus, trace visualization is most useful as an information-seeking layer that supports selective oversight, rather than as a replacement for conversational interaction.

\textbf{Automation vs. user control}
Participants also differed in how much they wanted to inspect and intervene in agent behavior. Some users primarily cared about the final outcome and trusted the agent to handle intermediate steps, whereas others frequently inspected the Trace Panel to maintain a stronger sense of control. This difference suggests that effective human--agent collaboration should accommodate a spectrum of oversight rather than assume a single monitoring style. As P7 observed, being able to ``ask about the current state more clearly, without having to stop it'' provides a lightweight alternative to interrupting the agent entirely.

\textbf{Adaptability vs. predictability}
The skill-evolution mechanism introduces another tension: agents should adapt to users' preferences without making their behavior opaque. Participants appreciated the automatic extraction of reusable knowledge from interaction histories. P4 described the distilled skill as ``really necessary'' because such knowledge previously required repeated human instructions. However, effective skill distillation requires transparency about the origin and applicability of learned behaviors. \name therefore exposes skill distillation histories and application conditions, allowing users to inspect learned behaviors rather than treating them as hidden adaptations.

\subsection{Implications for Human--Agent Collaboration}
Taken together, our findings suggest three design principles. First, \textbf{agent observability should be organized around information seeking rather than exhaustive monitoring}: hierarchical traces, summaries, and cross-panel linking allow users to quickly move from a high-level question to the relevant execution details. Second, \textbf{control should be granular and non-disruptive}: mechanisms such as trace forking allow users to revise a problematic decision while preserving useful preceding context, reducing the cost of restarting an interaction. Third, \textbf{agent learning should be grounded in observable interaction history}: distilling reusable skills from execution traces and user corrections turns individual interventions into persistent knowledge while keeping the provenance of that knowledge visible.

These principles point toward a broader shift in human--agent collaboration. Conventional coding-agent interaction largely follows a \emph{prompt--response} paradigm, in which users specify goals and react to completed outputs. Our findings suggest a complementary \emph{observe--intervene--learn} paradigm, where users can selectively inspect an agent's ongoing process, intervene at the level of execution context, and allow successful interventions to become reusable skills. In this model, the agent is not merely an autonomous executor that produces outputs; it becomes a continuously evolving collaborator whose behavior can be observed, corrected, and improved through interaction. \name provides one concrete example of this paradigm by connecting three previously separated layers---\emph{execution traces, user intervention, and persistent agent knowledge}---into a single interaction loop.

\section{Conclusion}

We presented \name, an interactive trace visualization and intervention system for human---coding agent collaboration.
We contribute both a structured trace visualization that helps users better observe and understand the behavior of agent execution and a trace-level interaction control that allows for more targeted agent interventions.
Additionally, the system features a trace-to-skill distillation mechanism that uses human interactions as feedback to automatically build or improve the reusable agent skills. 
The design of \name is validated through a quantitative evaluation of the online trace-to-skill algorithm, a user study assessing the trace visualization, and an expert interview to evaluate the entire system in real-world coding tasks.

\bibliographystyle{IEEEtran}
\bibliography{main}

\vspace{-2em}
\begin{IEEEbiography}[{\includegraphics[width=1in,height=1.25in,clip,keepaspectratio]{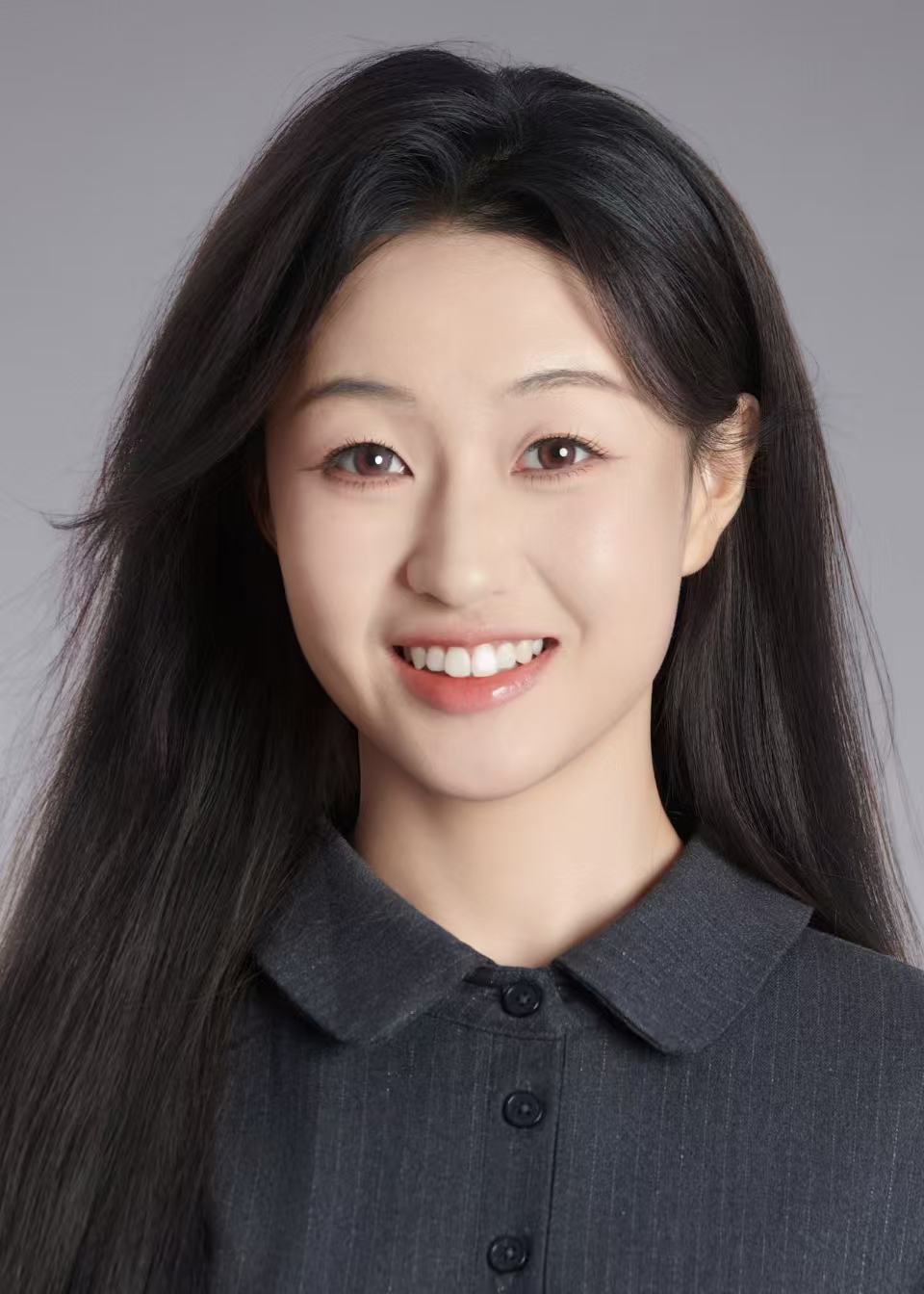}}]{Xiaoke Yan}
received her bachelor's degree from the Department of Software Engineering, Tongji University in 2024. She is currently pursuing her master's degree at the College of Design and Innovation, Tongji University, where she is a member of the Intelligent Design \& Innovation Lab (iDVx Lab). Her research interests include information visualization, AI-supported design, and human-computer interaction.
\end{IEEEbiography}

\vspace{-2em}
\begin{IEEEbiography}[{\includegraphics[width=1in,height=1.25in,clip,keepaspectratio]{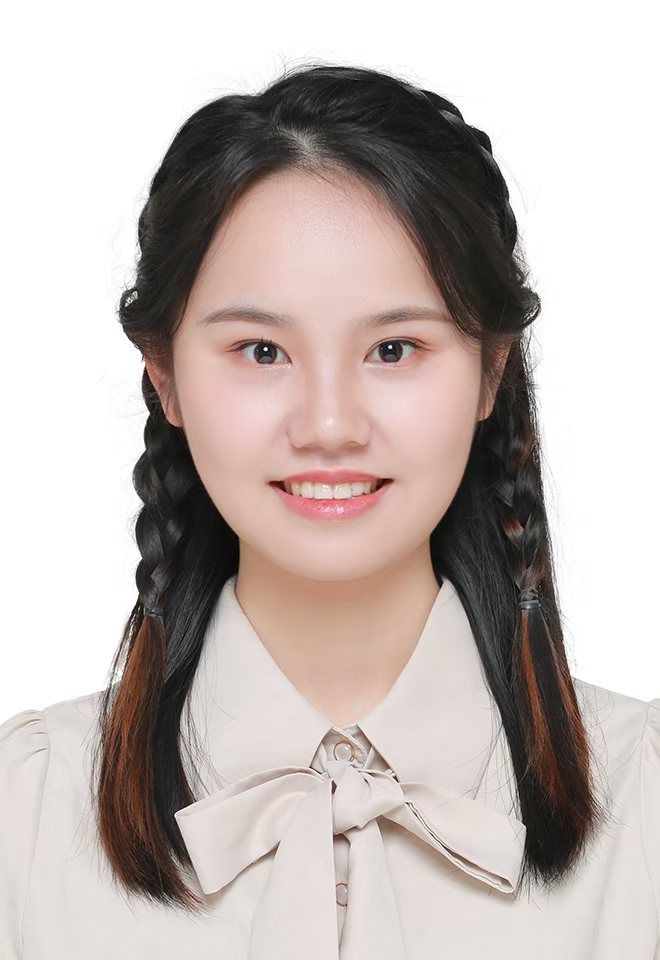}}]{Jiaying Lei}
received her bachelor's degree from the School of Mathematical Sciences, Tongji University, in 2024. She is currently a Ph.D. candidate jointly affiliated with the Shanghai Research Institute for Intelligent Autonomous Systems, Tongji University, and the Shanghai Innovation Institute. Her research interests include generative AI and human-computer interaction.
\end{IEEEbiography}

\vspace{-2em}
\begin{IEEEbiography}[{\includegraphics[width=1in,height=1.25in,clip,keepaspectratio]{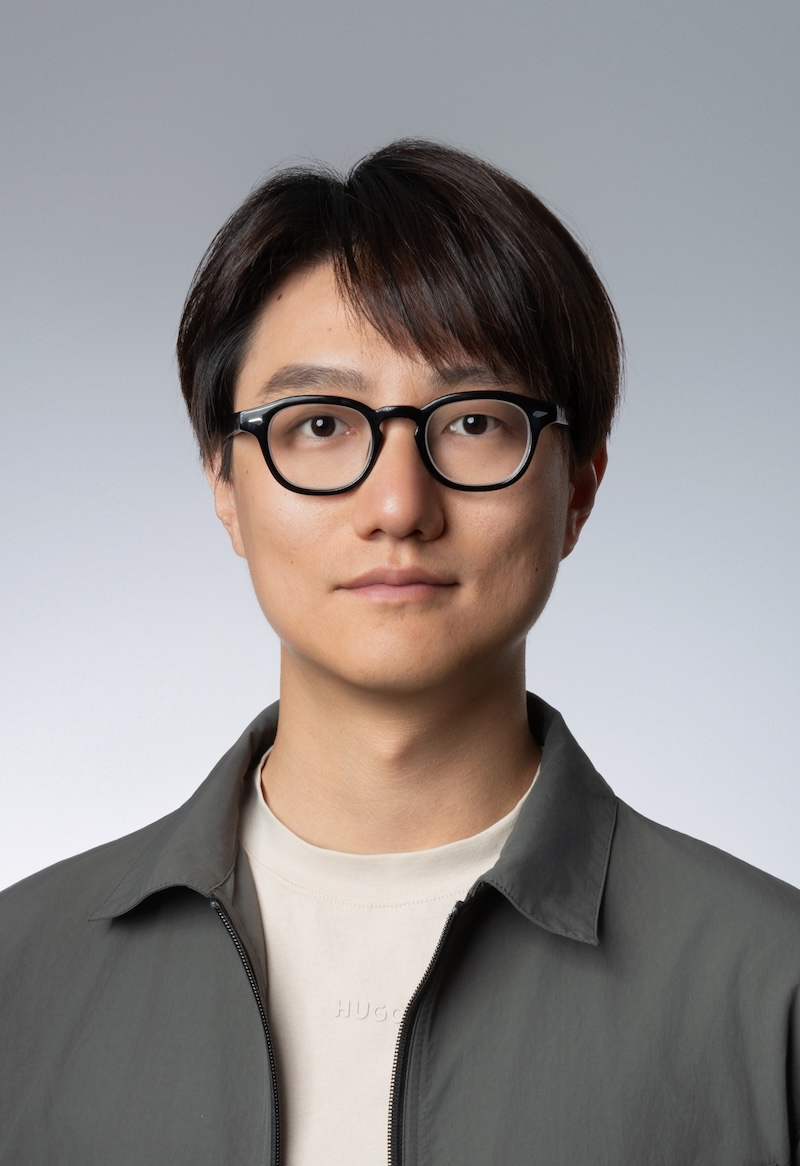}}]{Danqing Shi}
received his PhD degree from the College of Design and Innovation at Tongji University, China, in 2022. He is currently an assistant professor in the Department of Computer Science at Lund University, Sweden, where he leads the Lund HCI group. Before that, he was a postdoctoral research associate with the Department of Engineering at University of Cambridge, and an academic guest at ETH Zurich. His research interests include human-AI interaction, human-computer interaction, and visualization.
\end{IEEEbiography}

\vspace{-2em}
\begin{IEEEbiography}[{\includegraphics[width=1in,height=1.25in,clip,keepaspectratio]{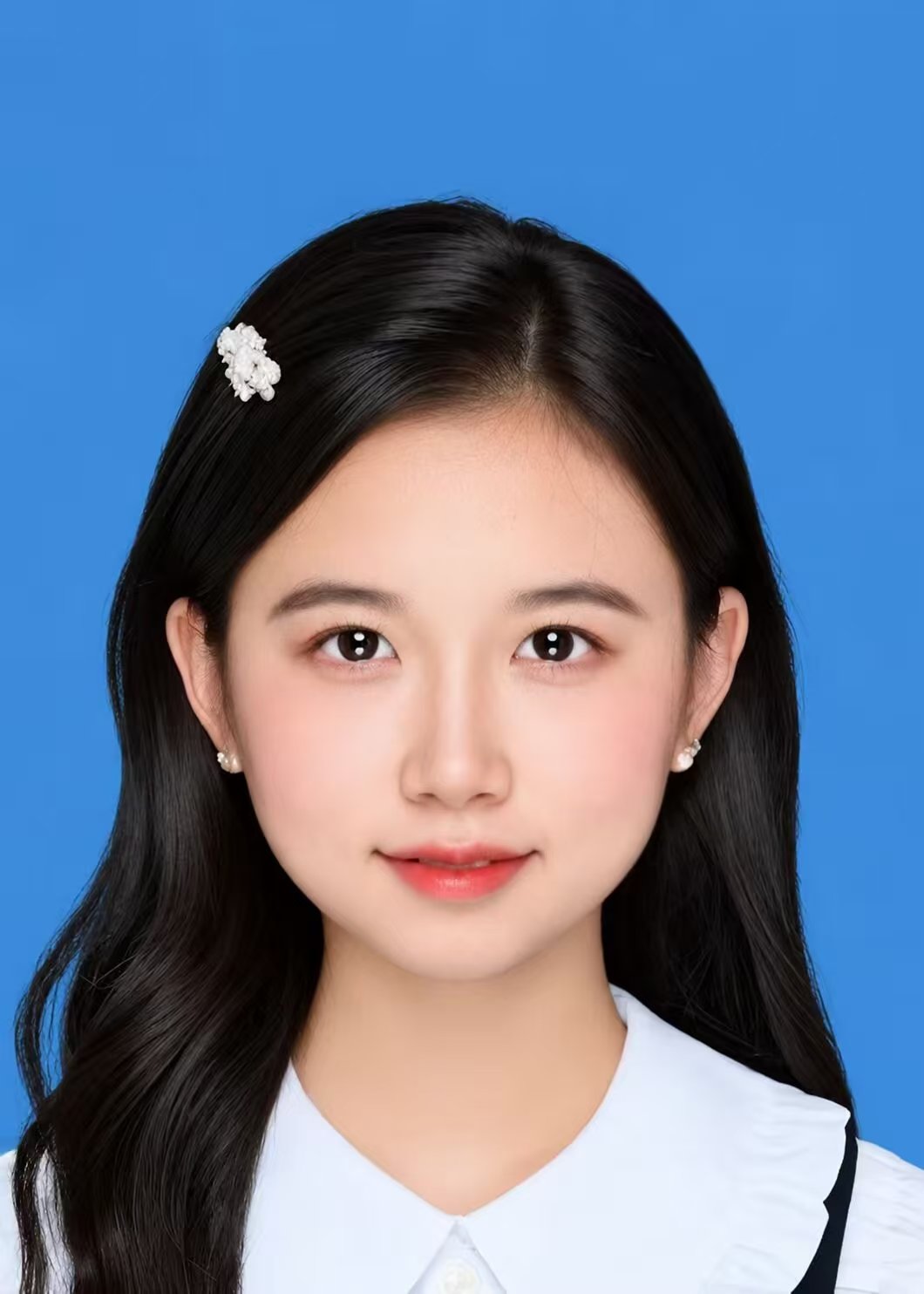}}]{Zichen Wang}
Wang Zichen, admitted to the School of Mathematical Sciences, Tongji University in 2023, is currently pursuing a Bachelor of Science in Mathematics and Applied Mathematics.
\end{IEEEbiography}

\vspace{-2em}
\begin{IEEEbiography}[{\includegraphics[width=1in,height=1.25in,clip,keepaspectratio]{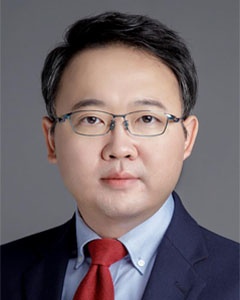}}]{Nan Cao}
received his Ph.D. degree in Computer Science and Engineering from the Hong Kong University of Science and Technology (HKUST), Hong Kong, China in 2012. He is currently a professor at Tongji University and the Vice Dean of the Tongji College of Design and Innovation. He also directs the Tongji Intelligent Big Data Visualization Lab (iDV$^x$ Lab) and conducts interdisciplinary research across multiple fields, including data visualization, human-computer interaction, and artificial intelligence. He was a research staff member at the IBM T.J. Watson Research Center, New York, NY, USA before joining the Tongji faculty in 2016.
\end{IEEEbiography}

\end{document}